\documentclass[sigconf,table,screen]{acmart} 

\AtBeginDocument{%
  \providecommand\BibTeX{{%
    \normalfont B\kern-0.5em{\scshape i\kern-0.25em b}\kern-0.8em\TeX}}}

\usepackage{subcaption}
\usepackage{enumitem}
\usepackage{microtype}
\usepackage{listings}
\usepackage{xspace}
\colorlet{punct}{red!60!black}
\definecolor{background}{HTML}{EEEEEE}
\definecolor{delim}{RGB}{20,105,176}
\colorlet{numb}{magenta!60!black}

\lstdefinelanguage{json}{
    basicstyle=\tiny\ttfamily,
    numberstyle=\scriptsize,
    numbersep=4pt, %8
    showstringspaces=false,
    breaklines=true,
    frame=lines,
    backgroundcolor=\color{background},
    literate=
     *
      {:}{{{\color{punct}{:}}}}{1}
      {,}{{{\color{punct}{,}}}}{1}
      {\{}{{{\color{delim}{\{}}}}{1}
      {\}}{{{\color{delim}{\}}}}}{1}
      {[}{{{\color{delim}{[}}}}{1}
      {]}{{{\color{delim}{]}}}}{1},
}
\lstdefinelanguage{shell}{
    basicstyle=\tiny\ttfamily,
    numberstyle=\scriptsize,
    numbersep=8pt,
    showstringspaces=false,
    breaklines=true,
    frame=lines,
    backgroundcolor=\color{background},
}

\theoremstyle{definition}
\newtheorem{definition}{Definition}[section]

\usepackage{tikz}
\usetikzlibrary{arrows,automata,positioning}

\def\Size{4pt}
\tikzset{
      folder/.pic={
        \filldraw[draw=folderborder,top color=folderbg!50,bottom color=folderbg]
          (-1.05*\Size,0.2\Size+5pt) rectangle ++(.75*\Size,-0.2\Size-5pt);  
        \filldraw[draw=folderborder,top color=folderbg!50,bottom color=folderbg]
          (-1.15*\Size,-\Size) rectangle (1.15*\Size,\Size);
      }
    }

\newcommand{\CHANGED}[1]{\textcolor{black}{#1}}
\definecolor{Gray}{gray}{0.9}
\graphicspath{{./}}

\usepackage{forest}
\definecolor{folderbg}{RGB}{124,166,198}
\definecolor{folderborder}{RGB}{110,144,169}

\usepackage{multirow}

\newcommand{\bb}[1]{\cellcolor{lightgray}#1}

\newcommand{\app}[1]{\textsc{#1}}

\newenvironment{revise}{\color{blue}}{\color{black}}
\newcommand{\toolurl}{\begin{revise}\url{https://bit.ly/30XLygW}\end{revise}}%https://github.com/dlicse1/DataLossDetector}\end{revise}}

\newcommand{\fullname}{Data Loss Detector\xspace}
\newcommand{\name}{DLD\xspace}
\newcommand{\alaric}{ALARic\xspace}

\newcommand{\rqa}{What is the $\epsilon$ that provides the best exploration?\xspace}
\newcommand{\rqb}{How effective is \name with data loss problems?\xspace}
\newcommand{\rqc}{Is \name more effective than state-of-the-art techniques?\xspace}
\newcommand{\rqd}{What is the tradeoff between the snapshot- and property-based oracles?\xspace}
\newcommand{\rqe}{Are data loss faults relevant to developers?\xspace}

\newcommand{\benchmarkfaults}{110\xspace}
\newcommand{\benchmarkfaultsNoLogin}{97\xspace}
\newcommand{\benchmarkApps}{48\xspace}
\newcommand{\benchmarkAppReleases}{54\xspace} %56

\newcommand{\revealedBenchmarkfaultsNoLogin}{73\xspace}
\newcommand{\revealedBenchmarkfaults}{83\xspace} 
\newcommand{\percentageRevealedBenchmarkfaults}{75\%\xspace}

\newcommand{\revealedBenchmarkfaultsNoLoginAlaric}{19\xspace}
\newcommand{\revealedBenchmarkfaultsNoLoginAlaricDLDgap}{3.8X\xspace} %73/19

\newcommand{\onlineFaults}{58\xspace}
\newcommand{\onlineFaultsRevealed}{35\xspace}
\newcommand{\percentageOnlineFaultsRevealed}{60\%\xspace}

\newcommand{\onlineFaultsRevealedAlaric}{8\xspace}
\newcommand{\onlineFaultsRevealedAlaricDLDGap}{4.4X\xspace} %35/8

\newcommand{\newFaultsRevealedNoLogin}{132\xspace} 
\newcommand{\newFaultsRevealed}{232\xspace} 
\newcommand{\newFaultsPerRelease}{4.3\xspace} %232/54

\newcommand{\newFaultsRevealedAlaric}{50\xspace}
\newcommand{\newFaultsRevealedAlaricDLDGap}{2.6X\xspace} %132/50

\newcommand{\totalFaultsRevealed}{350\xspace} %83+35+232
\newcommand{\missedBenchmarkfaults}{27\xspace} 
\newcommand{\onlineFaultsMissed}{23\xspace}
\newcommand{\totalMissedKnownFaults}{50\xspace} %23+27
\newcommand{\LowProbabilityMissedFaults}{41\xspace}
\newcommand{\EnvironmentMissedFaults}{4\xspace}
\newcommand{\NoActionMissedFaults}{5\xspace}

\newcommand{\ActivityNoLogin}{428\xspace}
\newcommand{\ActivityTotal}{731\xspace} 
\newcommand{\CoveredActivityNoLoginAvg}{65\%\xspace} 
\newcommand{\CoveredActivityNoLoginMax}{68\%\xspace} 
\newcommand{\CoveredActivityAvg}{62\%\xspace}
\newcommand{\CoveredActivityMax}{66\%\xspace}
\newcommand{\faultyActivityFoundNoLogin}{189\xspace}
\newcommand{\faultyActivityFound}{298\xspace}
\newcommand{\percentageFaultyActivityFound}{41\%\xspace} %298/731

\newcommand{\faultyActivityFoundNoLoginAlaric}{71\xspace}
\newcommand{\faultDiscoveryDifference}{2.7X\xspace} %189/71

\newcommand{\SpuriousViolationNoLogin}{11\xspace}
\newcommand{\AvgSpuriousViolationNoLogin}{0.26\xspace} % 11/42 
\newcommand{\SpuriousViolation}{14\xspace}
\newcommand{\AvgSpuriousViolation}{0.26\xspace} %14/54 

\newcommand{\SpuriousViolationAlaric}{31\xspace}
\newcommand{\AvgSpuriousViolationAlaric}{0.74\xspace} %31/42

\newcommand{\NumCrashesNoLogin}{26\xspace}
\newcommand{\NumCrashes}{38\xspace}

\newcommand{\dataLossDetectedByBoth}{73.1\%\xspace}
\newcommand{\dataLossDetectedByOneOracle}{26.9\%\xspace}
\newcommand{\dataLossDetectedByPropertyOracle}{90.9\%\xspace}
\newcommand{\dataLossDetectedBySnapshotOracle}{82.3\%\xspace}
\newcommand{\dataLossDetectedBySnapshotOracleOnly}{9.2\%\xspace}
 
\newcommand{\spuriousRotationOracle}{5.3\%\xspace}
\newcommand{\spuriousViolationsByPropertyOracle}{0.1\%\xspace}
\newcommand{\spuriousViolationsBySnapshotOracle}{21.1\%\xspace}
\newcommand{\spuriousViolationsByBoth}{73.6\%\xspace}

\newcommand{\feedbackReceived}{98\xspace}
\newcommand{\bugReportsConfirmed}{88\xspace}
\newcommand{\percentageBugReportsConfirmed}{90\%\xspace}
\newcommand{\bugReportsNotConfirmed}{10\xspace}
\newcommand{\bugReportsNotRelevant}{33\xspace}

\acmConference[ISSTA '20]{Proceedings of the 29th ACM SIGSOFT International Symposium on Software Testing and Analysis}{July 18--22, 2020}{Virtual Event, USA}
\acmBooktitle{Proceedings of the 29th ACM SIGSOFT International Symposium on Software Testing and Analysis (ISSTA '20), July 18--22, 2020, Virtual Event, USA}

\begin{document}

\title{\fullname: Automatically Revealing Data Loss Bugs in Android Apps}

\author{Oliviero Riganelli}
\email{oliviero.riganelli@unimib.it}
\affiliation{%
  \institution{University of Milano - Bicocca}
  \city{Milan}
  \country{Italy}
  \postcode{20126}
}

\author{Simone Paolo Mottadelli}
\email{s.mottadelli2@campus.unimib.it}
\affiliation{%
  \institution{University of Milano - Bicocca}
  \city{Milan}
  \country{Italy}
  \postcode{20126}
}

\author{Claudio Rota}
\email{c.rota30@campus.unimib.it}
\affiliation{%
  \institution{University of Milano - Bicocca}
  \city{Milan}
  \country{Italy}
  \postcode{20126}
}

\author{Daniela Micucci}
\email{daniela.micucci@unimib.it}
\affiliation{%
  \institution{University of Milano - Bicocca}
  \city{Milan}
  \country{Italy}
  \postcode{20126}
}

\author{Leonardo Mariani}
\email{leonardo.mariani@unimib.it}
\orcid{0000-0001-9527-7042}
\affiliation{%
  \institution{University of Milano - Bicocca}
  \city{Milan}
  \country{Italy}
  \postcode{20126}
}

%
%\author{Lars Th{\o}rv{\"a}ld}
%\affiliation{%
%  \institution{The Th{\o}rv{\"a}ld Group}
%  \streetaddress{1 Th{\o}rv{\"a}ld Circle}
%\email{larst@affiliation.org}
%\affiliation{%
%  \institution{Inria Paris-Rocquencourt}
%\affiliation{%
% \institution{Rajiv Gandhi University}
%\affiliation{%
%  \institution{Tsinghua University}
%  \streetaddress{30 Shuangqing Rd}
%\affiliation{%
%  \institution{Palmer Research Laboratories}
%  \streetaddress{8600 Datapoint Drive}
%\affiliation{\institution{The Th{\o}rv{\"a}ld Group}}
%\email{jsmith@affiliation.org}
%\affiliation{\institution{The Kumquat Consortium}}
%\email{jpkumquat@consortium.net}

\renewcommand{\shortauthors}{Oliviero Riganelli, Simone Paolo Mottadelli, Claudio Rota, Daniela Micucci and Leonardo Mariani}

\begin{abstract}

Android apps must work correctly even if their execution is interrupted by external events. For instance, an app must work properly even if a phone call is received, or after its layout is redrawn because the smartphone has been rotated. Since these events may require destroying, when the execution is interrupted, and recreating, when the execution is resumed, the foreground activity of the app, the only way to prevent the loss of state information is to save and restore it. This behavior must be explicitly implemented by app developers, who often miss to implement it properly, releasing apps affected by \emph{data loss} problems, that is, apps that may lose state information when their execution is interrupted.  

Although several techniques can be used to automatically generate test cases for Android apps, the obtained test cases seldom include the interactions and the checks necessary to exercise and reveal data loss faults. To address this problem, this paper presents \emph{\fullname} (\name), a test case generation technique that integrates an exploration strategy, data-loss-revealing actions, and two customized oracle strategies for the detection of data loss failures. 

\name revealed 
%has been able to reveal 
\percentageRevealedBenchmarkfaults of the faults in a benchmark of \benchmarkAppReleases Android app releases affected by \benchmarkfaults known data loss faults, and also revealed unknown data loss problems, outperforming competing approaches.

\end{abstract}

\begin{CCSXML}
<ccs2012>
<concept>
<concept_id>10011007.10011074.10011099.10011102.10011103</concept_id>
<concept_desc>Software and its engineering~Software testing and debugging</concept_desc>
<concept_significance>500</concept_significance>
</concept>
</ccs2012>
\end{CCSXML}

\ccsdesc[500]{Software and its engineering~Software testing and debugging}
\keywords{Android, data loss, test case generation, validation, mobile apps.}


\maketitle

% !TEX root =  main.tex
\section{Introduction}\label{sec:introduction}
In the last decade, mobile apps have increasingly gained importance and popularity. Recent studies revealed that people spend more than 3h per day on their smartphones on average~\cite{RescueTime2019} and that 90\% of this time is typically devoted to the use of mobile apps~\cite{Mobiloud2018}. %interactions with mobile apps~\cite{Mobiloud2018}.  
Indeed, people use mobile apps to perform a huge variety of tasks, including reading e-mails, listening to music, making orders and payments, and playing games. 

Among the available ecosystems for the distribution of mobile apps, the Android ecosystem is the largest and most used one: its market share is almost 75\%~\cite{Statcounter2019} and its official store, the Google Play Store, includes almost 3.0 millions of apps~\cite{GraphWithAndroidAppsNumberInGooglePlayStore}. %Indeed, Android apps are the most popular mobile apps.

% for fun and for work, thus performing activities such as reading e-mails, surfing the net, listening to music, making payments, and so on. %They use apps both for fun and for work, entrusting them with confidential data such as work e-mails and banking passwords. 

%The Android ecosystem is the largest ecosystem of mobile apps: it has a market share higher than 75\%~\cite{Statcounter2019} and more than 2.5 millions of apps are available in the Google Play Store~\cite{GraphWithAndroidAppsNumberInGooglePlayStore}. 

Android apps consist of components, such as activities, fragments, and services, whose behavior must comply with well-defined lifecycles~\cite{AndroidLifeCycle,AndroidFragmentLifeCycle,Service}. For instance, activities can be in states such as created, paused, resumed, and stopped, and transitions between these states produce \emph{callbacks} % (e.g., to \texttt{onCreate()}, \texttt{onStart()}, \texttt{onPause()}, \texttt{onResume()}, \texttt{onStop()}) 
that must be handled by the activities.	

Interestingly, some of these callbacks might be particularly tricky to implement. This is the case of the callbacks produced by \emph{stop-start} events, which are system events that may force the destruction and then the (re-)instantiation of a running activity.
%an activity to first enter the \textit{stopped} state, which causes the destruction of the activity, and then the \textit{resumed} state, which causes the (re-)instantiation of the same activity. 
Stop-start events occur every time the execution of an app is stopped and then resumed. Typical cases include answering a phone call, switching between apps, and rotating the smartphone to change its layout.

When a stop-start event occurs, the difficult task for the app is to handle the destruction of the current activity in a way that is later possible to resume the execution at the same point it was interrupted. This is done by saving the values of all the relevant state variables before the activity is destroyed, and retrieving these values when the execution is resumed. With the exception of some specific cases (e.g., the widgets with a non-empty \textit{android:id} property), it is a responsibility of the developer to implement this behavior. In particular, developers have to implement both the logic necessary to save the state of an activity in the {\small \texttt{onSaveInstanceState}()} callback method and the logic to resume its state in the {\small \texttt{onRestoreInstanceState}()} callback method~\cite{SavingUIStates}. Unfortunately, this implementation might be wrong and might introduce misbehaviors in the apps~\cite{Hu:2011:AGT}. 

When a stop-start event is not properly handled, the Android app is said to be affected by a \emph{data loss} fault, that is, a fault that causes one or more state variables to lose their values. Data loss faults may affect the correctness of the apps in many different ways. In the best cases, they force the users to enter again inputs that had already been entered, deteriorating the quality of the user experience. In the worst cases, they generate activities with an inconsistent state, which causes the apps to produce wrong outputs or even crashes.

Data loss faults can be extremely pervasive. Adamsen \emph{et al.}~\cite{SystematicExecutionOfAndroidTestSuitesInAdverseConditions} considered the execution of system events jointly with test suites and reported that all the four apps used in their evaluation were affected by data loss faults. Riganelli \emph{et al.}~\cite{ABenchmarkOfDataLossBugsForAndroidApps} analyzed 428 Android apps and found that at least 82 (19.6\%) of the apps were affected by data loss faults. %, some of which required more than 1 year to be fixed. 
Finally, Amalfitano  \emph{et al.}~\cite{WhyDoesTheOrientationChangeMessUpMyAndroidApplication} studied 68 open source apps reporting data loss faults in 60 of them (88.2\%).

Test case generation techniques could be potentially used to reveal data loss faults. Indeed, a number of automatic test case generation techniques are available for Android apps. For instance, Monkey can generate test cases randomly~\cite{Monkey}, A$^3$E can generate tests systematically using a depth-first strategy~\cite{A3E}, DroidBot~\cite{Droidbot} and Stoat~\cite{Stoat} exploit a model-based approach, and Sapienz uses evolutionary algorithms~\cite{Sapienz}. While these approaches are able to reveal several interesting faults, they are \emph{ineffective} against data loss problems for two reasons: (i) they do not include operations that cause stop-start events, and (ii) they are not equipped with oracles strong enough to detect non-crashing data loss failures. 

Some techniques have been designed to extend the fault discovery capability of existing test suites to data loss problems. For instance, Thor systematically injects neutral event sequences, including stop-start events, into existing test cases to augment their failure detection capability~\cite{SystematicExecutionOfAndroidTestSuitesInAdverseConditions}. Quantum behaves similarly but it starts from a GUI model of the app~\cite{Razieh:OraclesUserInteraction:ICST:2014}. Although useful, their applicability is limited to apps equipped with comprehensive test suites or GUI models. \alaric randomly generates test cases that include stop-start events and detects data loss problems by comparing the state of the app under test before and after a stop-start event is generated~\cite{alaric}. \alaric can successfully reveal data loss faults, but the adopted \CHANGED{exploration} strategy and oracles have limited effectiveness, as reported in our evaluation.

In this paper, we present \emph{\fullname} (\name), an automatic testing technique that can reveal data loss problems in Android apps. \name integrates three capabilities to effectively reveal data loss problems: (i) a \emph{biased model-based} exploration strategy that steers the exploration towards (new) app states that may be affected by data loss problems, (ii) \emph{data-loss-revealing actions} that increase the likelihood to expose data loss problems, and (iii) two state-based oracles, a \emph{snapshot-based} oracle and a \emph{property-based} oracle, that have a high data loss detection accuracy, especially if used jointly.

\CHANGED{Compared to Thor~\cite{SystematicExecutionOfAndroidTestSuitesInAdverseConditions} and Quantum~\cite{Razieh:OraclesUserInteraction:ICST:2014}, DLD does not require a pre-existing test suite or a pre-existing GUI model, neither requires an initial ripping phase, but iteratively and continuously generates test cases according to the allocated time budget. %While discovering data loss faults, DLD builds a model that is similar to the one required by Quantum. However, DLD uses an abstraction that supports the detection of data loss problems on GUI elements not immediately visible (e.g., that can be visualized scrolling down or up), which are not detectable with Quantum. 
Finally, the biased model-based exploration and the data-loss-revealing actions exploited in DLD allow a more effective data loss detection than the strategy implemented in Alaric~\cite{alaric}.}

We empirically evaluated \name using the benchmark by Riganelli \emph{et al.}~\cite{ABenchmarkOfDataLossBugsForAndroidApps}, which includes \benchmarkfaults data loss problems affecting \benchmarkAppReleases app releases. \name automatically detected \revealedBenchmarkfaults of the \benchmarkfaults (\percentageRevealedBenchmarkfaults) data loss problems. \name also revealed \onlineFaultsRevealed data loss faults that were not part of the benchmark, but were reported online in bug reports, and \newFaultsRevealed previously unknown data loss faults. \CHANGED{Overall \name revealed three times the data loss faults revealed  by competing approaches.} %The empirical comparison between the snapshot-based and property-based oracles revealed interesting complementarities. 
We finally submitted the data loss faults that still affect apps nowadays online to app developers who positively reacted to our bug reports.   

In a nutshell, this paper makes the following contributions.
\begin{itemize}[leftmargin=*]
\item It describes an exploration strategy, data-loss-revealing actions, and two state-based oracles that can be incorporated in testing techniques to reveal data loss problems,
\item It delivers the \fullname (\name) 
 technique, which is implemented as an extension of the DroidBot~\cite{Droidbot} test case generation tool for Android,
\item It reports the \CHANGED{largest empirical evaluation about data loss detection available so far}, considering hundreds of data loss faults, % and  test case generation techniques, demonstrating the high effectiveness of \name, also in comparison to \alaric.
\item It delivers the tool and the experimental material freely available online at the following url: \toolurl %\footnote{We did our best to anonymize all the material, but we cannot exclude that working with the online material our identity might be incidentally revealed.}.
\end{itemize}

This paper is organized as follows. Section~\ref{sec:dataloss} discusses data loss faults and exemplifies typical failures that can be experienced with Android apps. 
Section~\ref{sec:approach} describes the main technical contributions of this paper, including the biased model-based exploration strategy, the data-loss-revealing actions, and the automatic oracles. 
Section~\ref{sec:evaluation} reports empirical results. Section~\ref{sec:related} discusses related work. Finally, Section~\ref{sec:conclusion} provides final remarks.  

%The article is structured as follows: section \ref{chapter:Data loss failure in Android apps} characterizes the \textit{Data Loss} problem and its causes, showing different examples from real world apps; chapter \ref{chapter:State of the art} discusses the state of the art of Android testing, focusing itself on both the available state-of-the-art test case generation tools for Android apps and how \textit{Data Loss} problems are dealt with nowadays; chapter \ref{chapter:Development} shows the steps followed for the integration of the novel technique into an existing tool, providing an overview of its functioning and describing its implementation; chapter \ref{chapter:Evaluation} reports the results of the evaluation of this technique, showing its abilities in both exploring the apps under test and detecting \textit{Data Loss} problems, and, finally, chapter \ref{chapter:Conclusions} concludes this thesis briefly summarizing the contents of this work and mentioning future improvements.

% !TEX root =  main.tex
\section{Data Loss Faults}\label{sec:dataloss}

In this section we describe data loss faults and the Android components that can be affected by these faults: activities and fragments.

An Android \emph{activity} is a component that implements a screen of an app and the logic to handle that screen. To partition a screen into smaller units, activities can contain a number of \emph{fragments}, each one containing both some graphical elements and the logic to handle them. Both activities and fragments have their own lifecycle~\cite{AndroidLifeCycle,AndroidFragmentLifeCycle}.

% !TEX root =  main.tex
\begin{figure*}[hbt!]
\begin{subfigure}[h]{0.49\linewidth}
    \begin{subfigure}[h]{0.39\linewidth}
        \includegraphics[width=\linewidth]{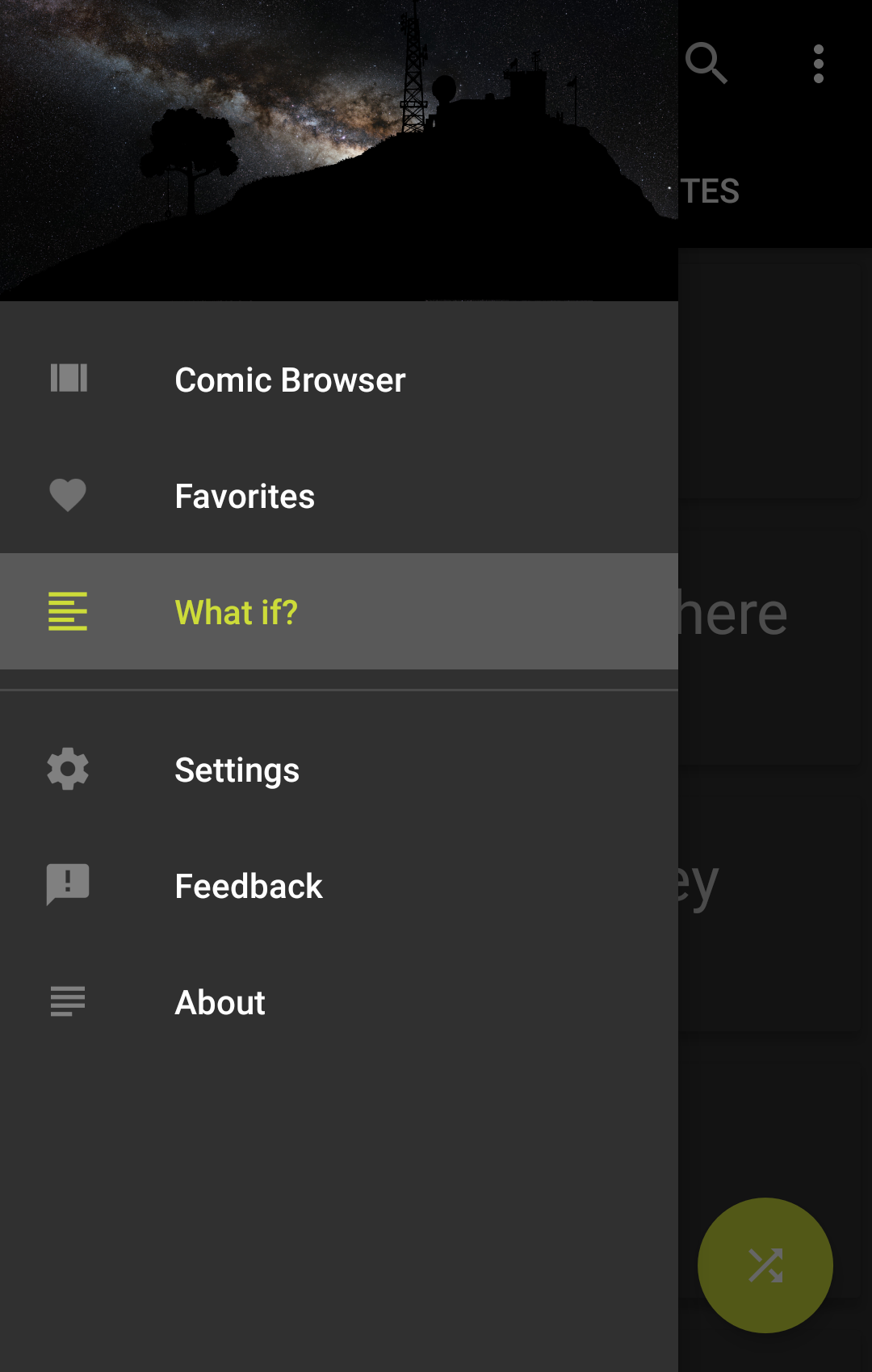}
        %\caption{Before a \textit{stop-start} event}
    \end{subfigure}
       \begin{subfigure}[h]{0.6\linewidth}
        \includegraphics[width=\linewidth]{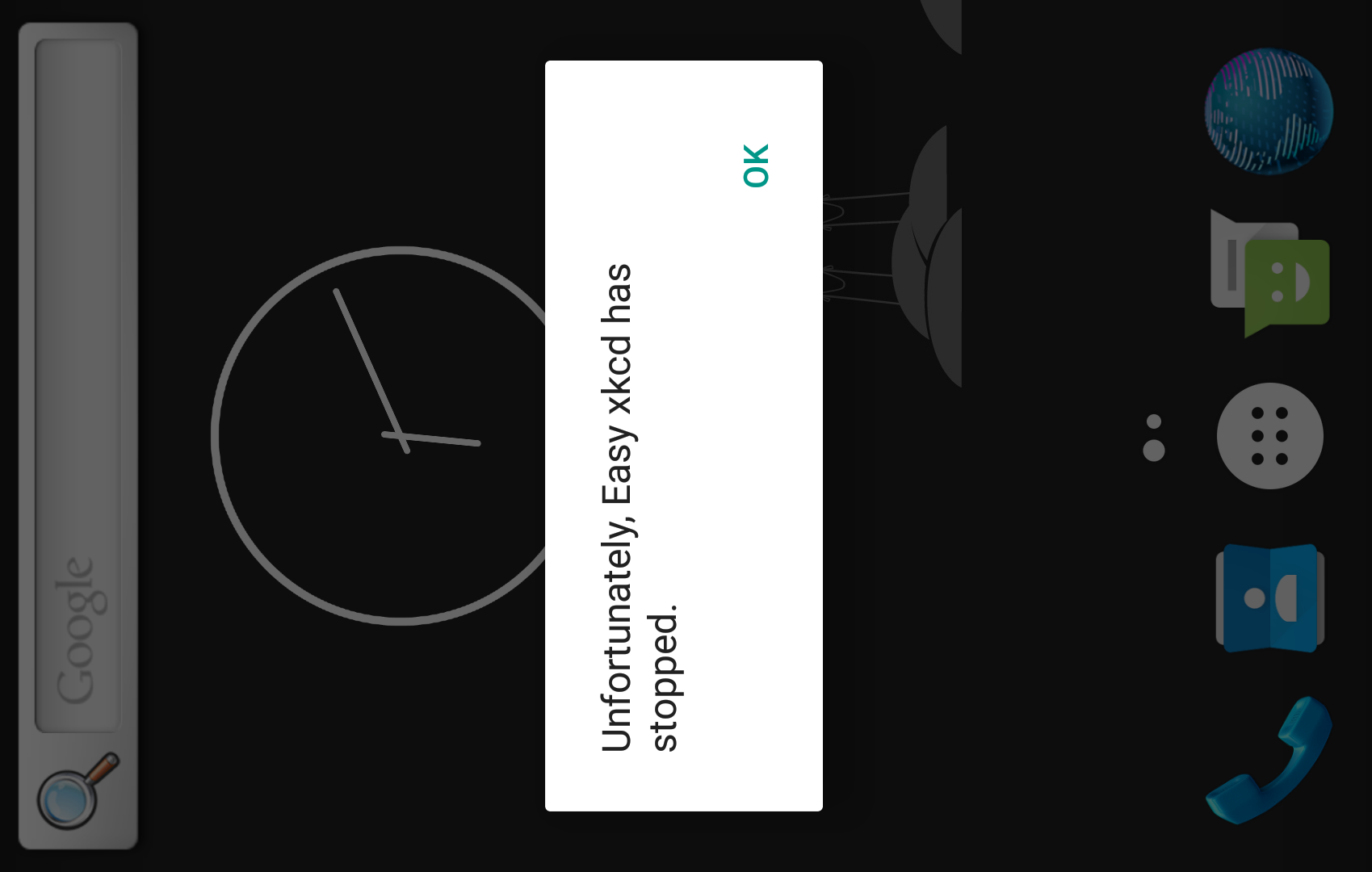}
       %\caption{After a \textit{stop-start} event}
    \end{subfigure} 
     \caption{Application crash in \app{Easy xkcd} v6.0.4}
     \label{fig:dl0.39-crash}
\end{subfigure} 
\begin{subfigure}[h]{0.49\linewidth}
  \begin{subfigure}[h]{0.39\linewidth}
        \includegraphics[width=\linewidth]{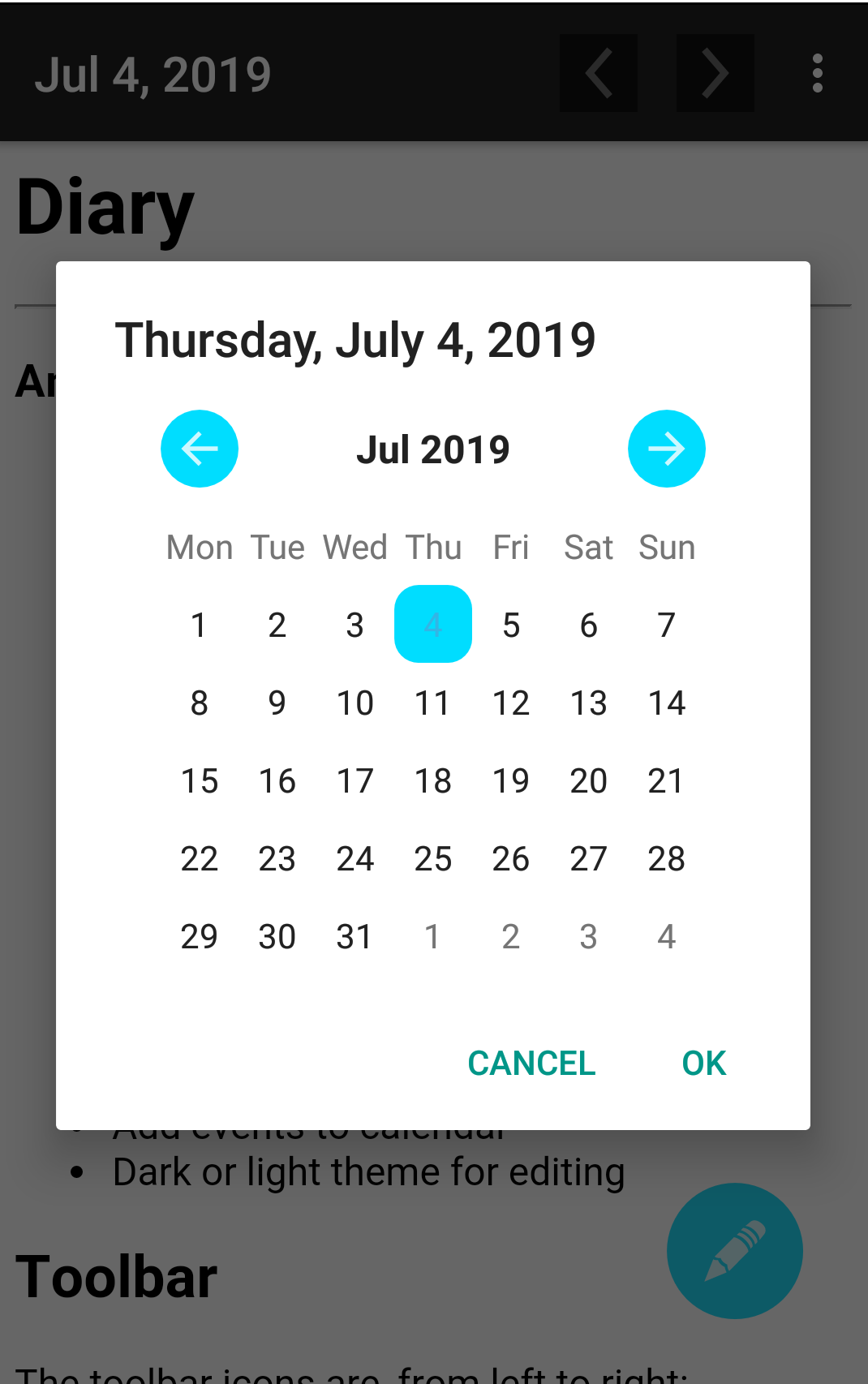}
        %\caption{Before a \textit{stop-start} event}
    \end{subfigure}
       \begin{subfigure}[h]{0.6\linewidth}
        \includegraphics[width=\linewidth]{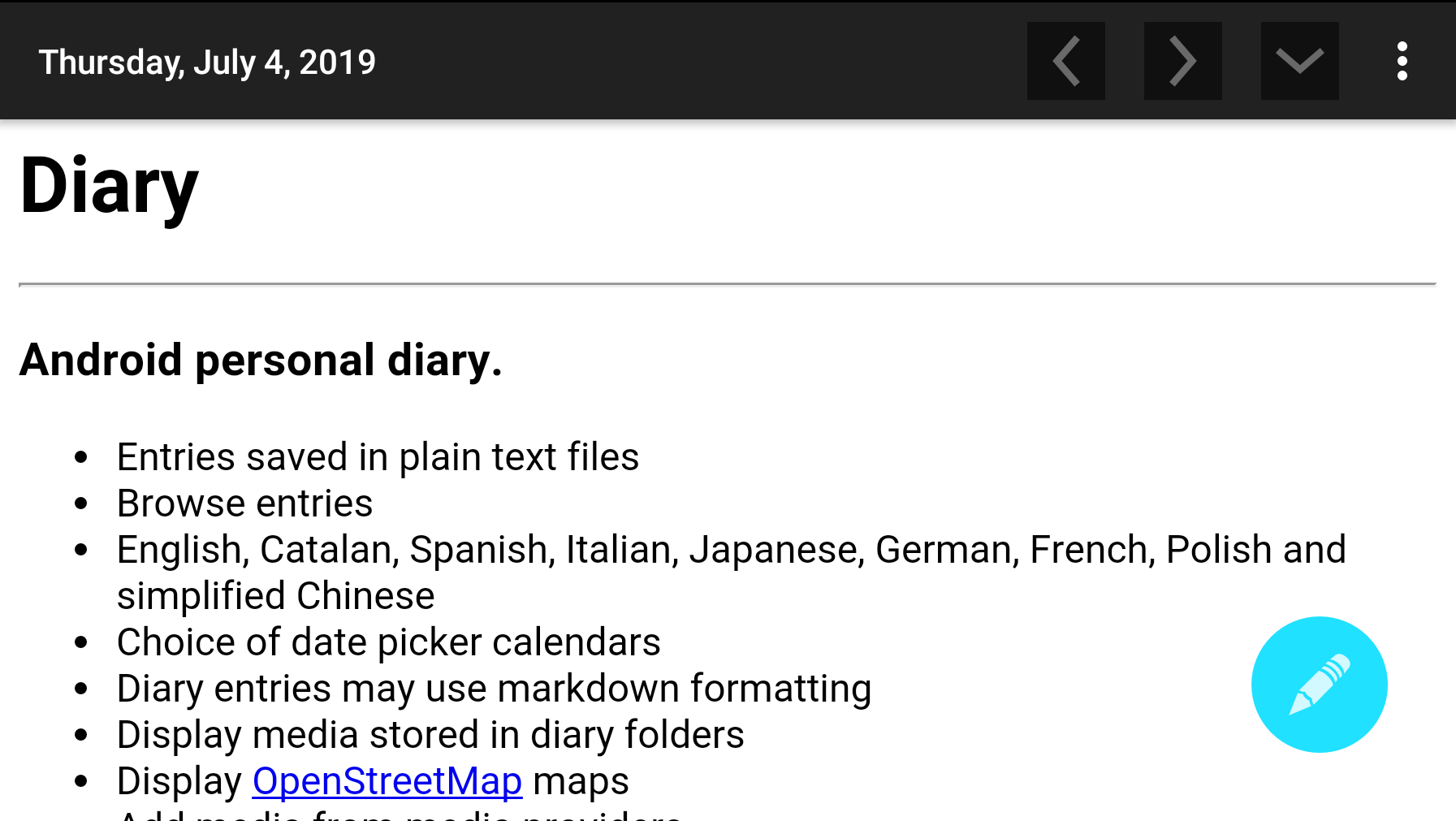}
        %\caption{After a \textit{stop-start} event}
    \end{subfigure}
	\caption{Disappearance of a \emph{Dialog} in \app{Diary} v1.26}
     \label{fig:dl-disappearance}
\end{subfigure} 
    \par\medskip
    
\begin{subfigure}[h]{0.49\linewidth}
	\begin{subfigure}[h]{0.39\linewidth}
        		\includegraphics[width=\linewidth]{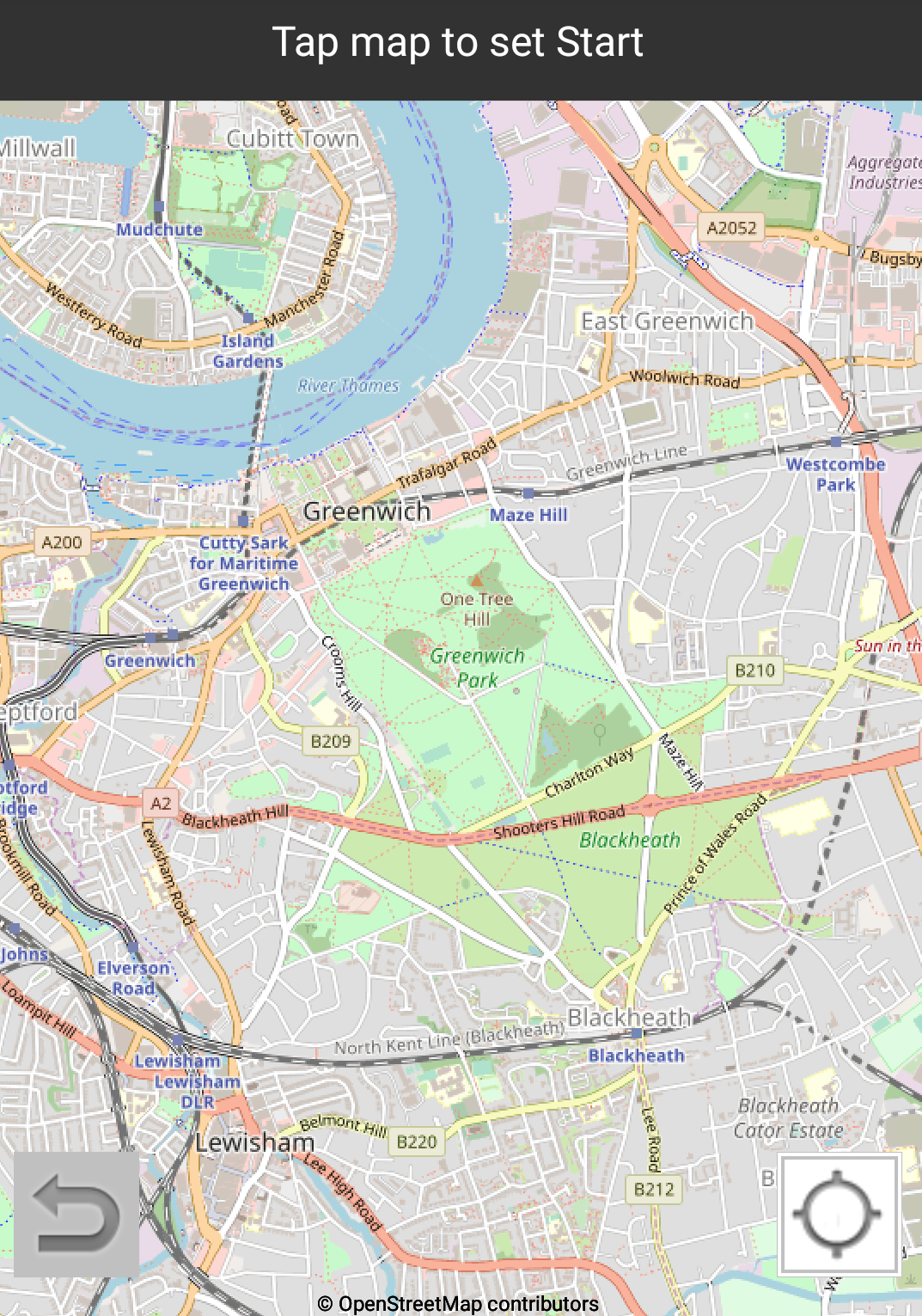}
		%\caption{Before a \textit{stop-start} event}
	\end{subfigure}
	\begin{subfigure}[h]{0.6\linewidth}
        		\includegraphics[width=\linewidth]{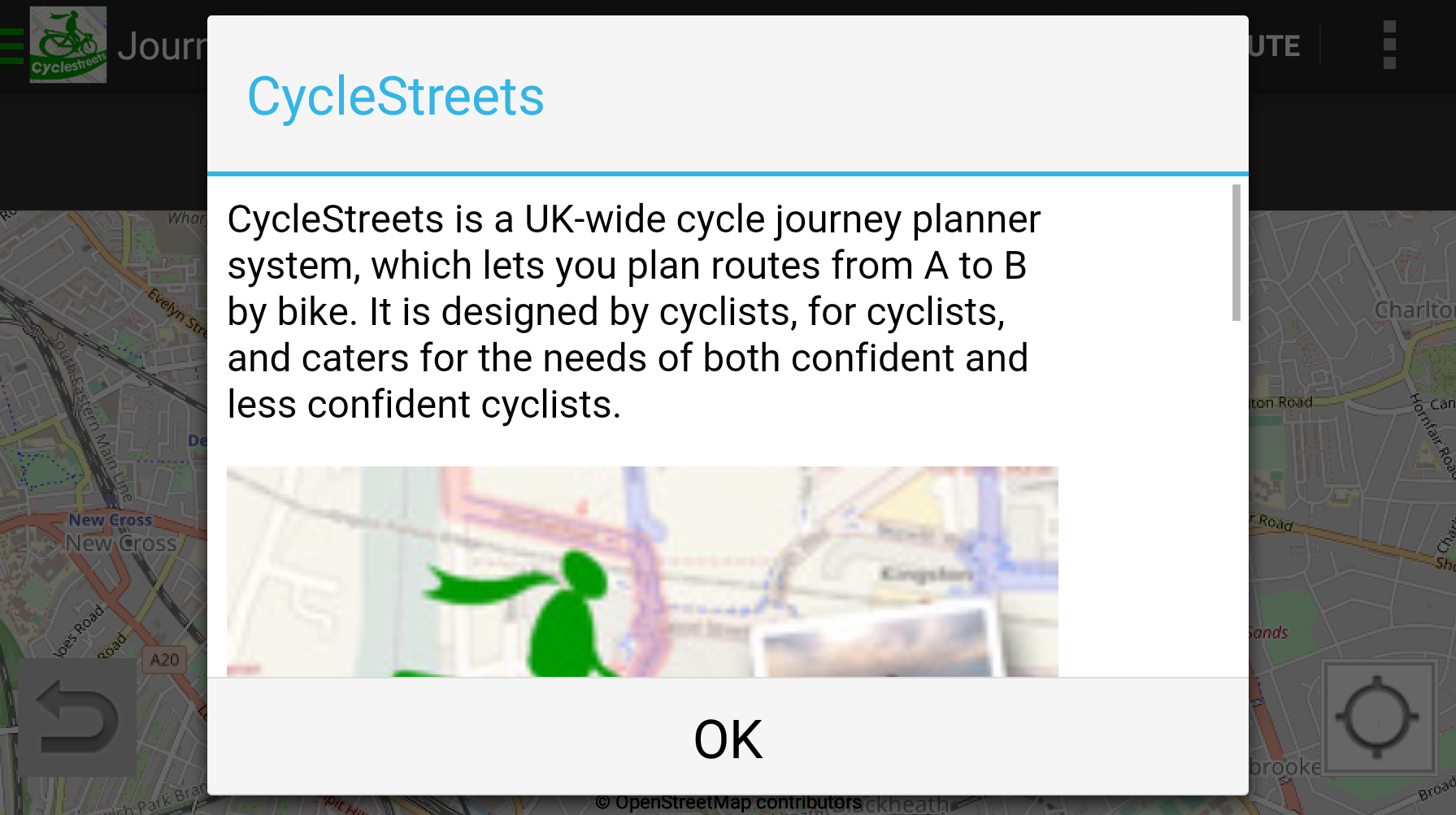}
        		%\caption{After a \textit{stop-start} event}
	\end{subfigure}
	  \caption{Appearance of a \textit{Dialog} in \app{CycleStreets} v3.5}
     	\label{fig:dl-appearance}
\end{subfigure} 
\begin{subfigure}[h]{0.49\linewidth}
	\begin{subfigure}[h]{0.39\linewidth}
        		\includegraphics[width=\linewidth]{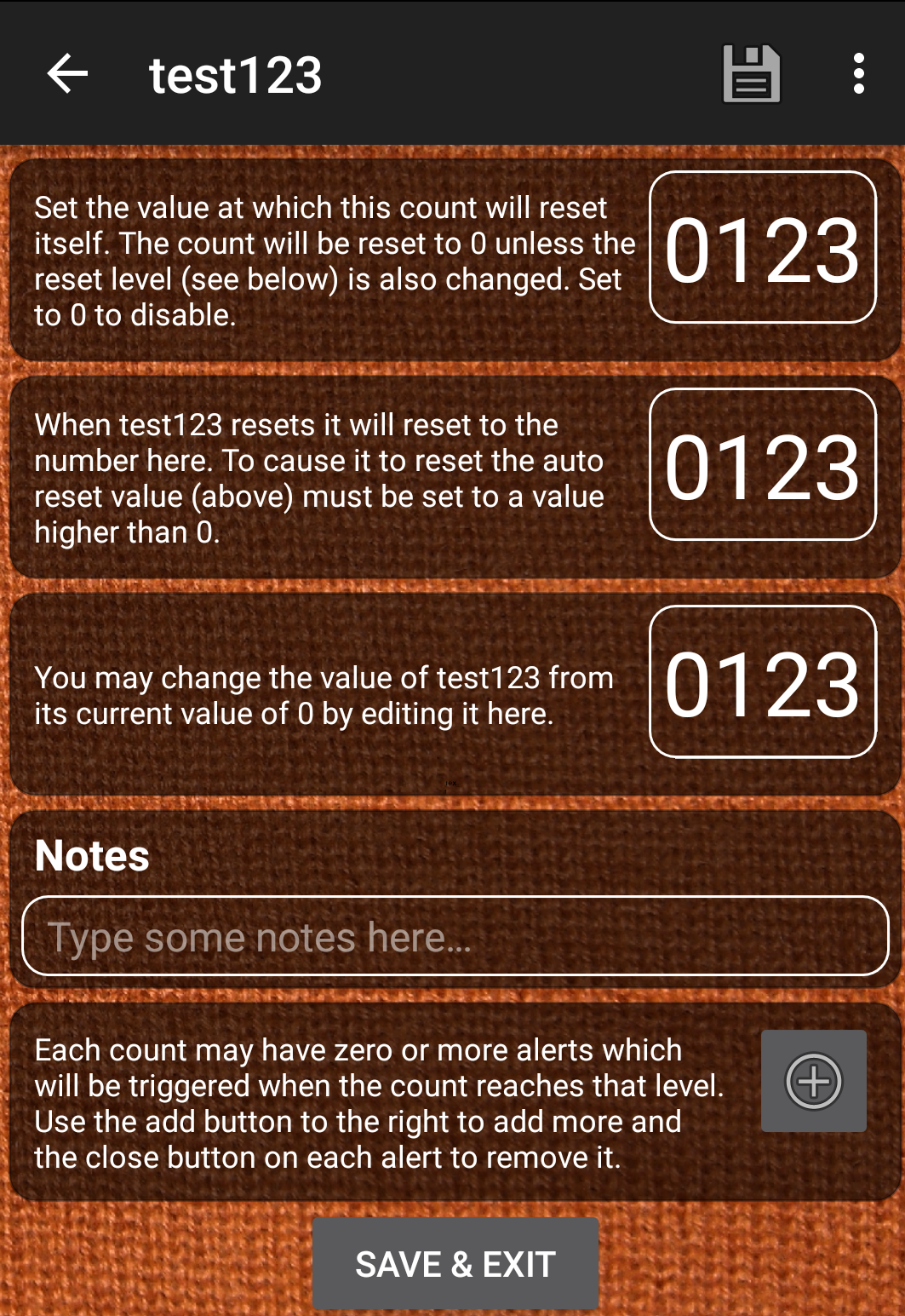}
        		%\caption{Before a \textit{stop-start} event}
    	\end{subfigure}
        \begin{subfigure}[h]{0.6\linewidth}
        		\includegraphics[width=\linewidth]{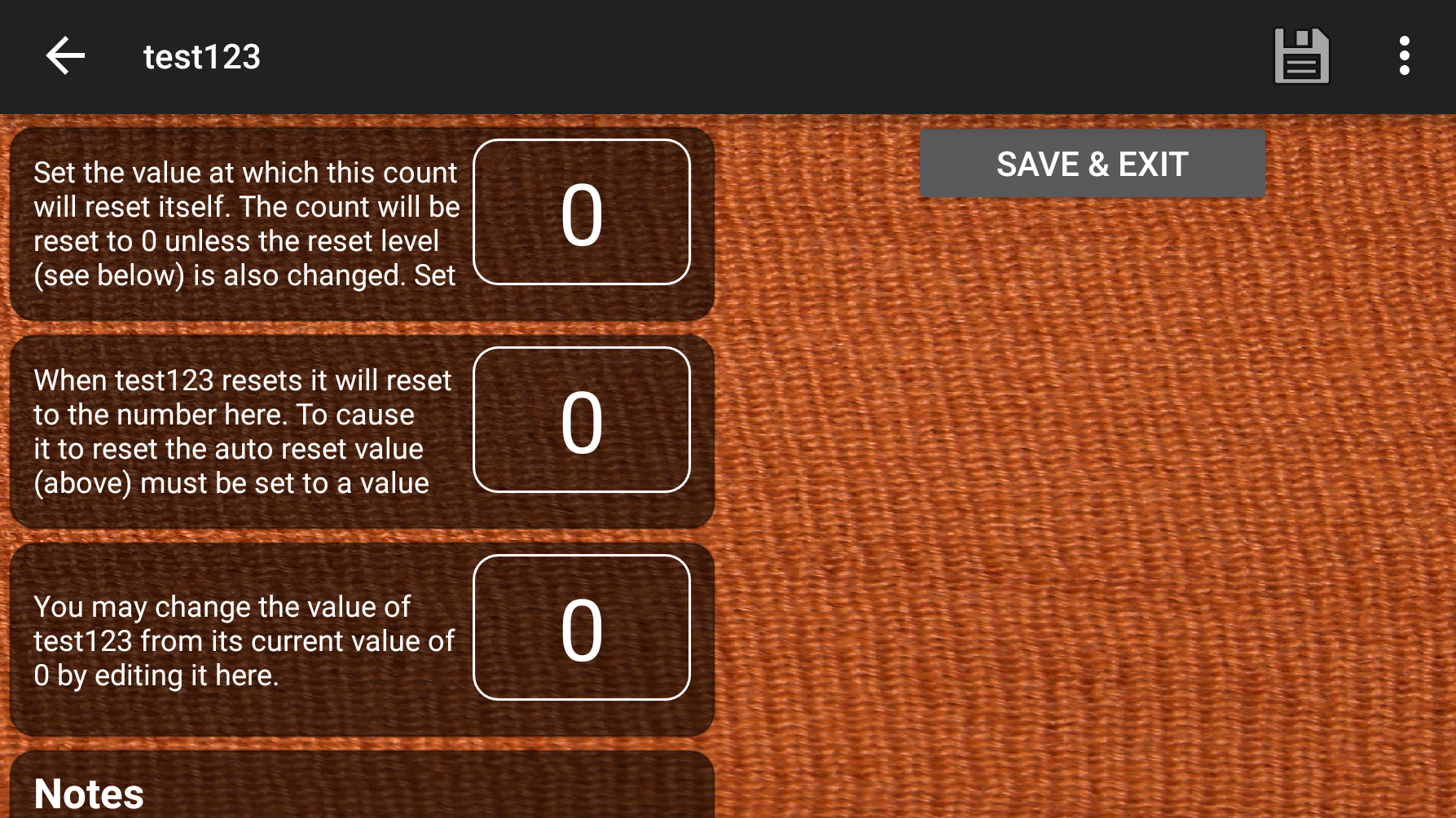}
        		%\caption{After a \textit{stop-start} event}
     	\end{subfigure}
	\caption{Change of  \textit{EditText} values in \app{BeeCount} v2.7.4}
	\label{fig:dl-wrong}
	\end{subfigure} 
	\par\medskip
\begin{subfigure}[h]{\linewidth}
\begin{center}
	\begin{subfigure}[h]{\linewidth}
        		\includegraphics[width=\linewidth]{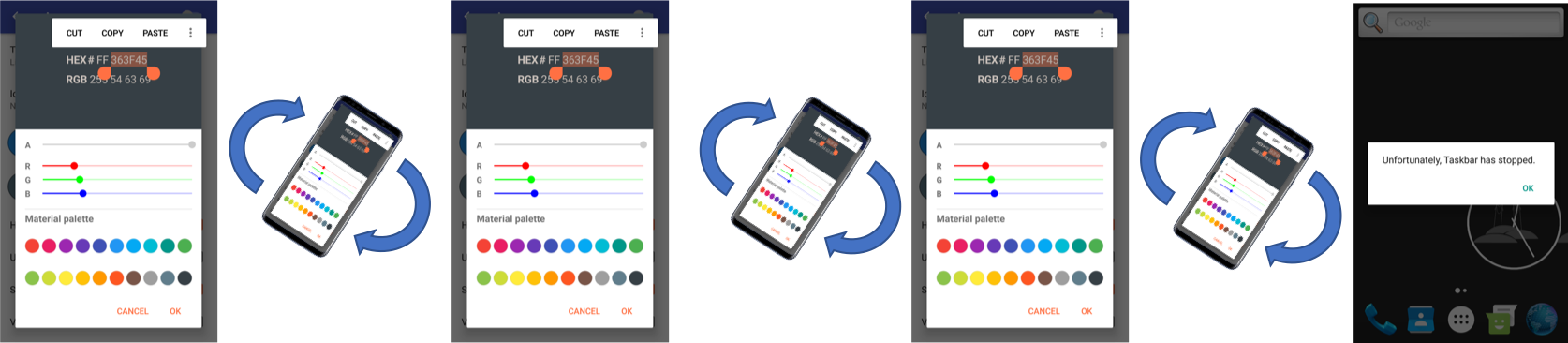}
		%\caption{Before a \textit{stop-start} event}
	\end{subfigure}
\end{center}	
	  \caption{Loss of the internal state in \app{Taskbar} v3.0.3}
     	\label{fig:stateLost}
\end{subfigure} 

\caption{Examples of data loss failures after a stop-start event.}
\label{fig:crash}
\end{figure*}

Specific sequences of system events may have a direct impact on the lifecycle of activities and fragments. %In particular,

\vspace{-0.1cm}
\begin{definition}
A \textit{stop-start} event is a sequence of system events that may cause a running activity (or fragment) to be destroyed and then recreated\footnote{In the rest of the paper we refer only to activities for simplicity, but all the concepts apply to both activities and fragments.}.
\end{definition}
%\begin{definition}
%A \textit{stop-start} event is a sequence of system events that may cause a running activity (or fragment) to enter the stopped state, which implies the destruction of the activity (or of the fragments), followed by the resumed state, which causes the recreation of the stopped activity (or of the stopped fragments)\footnote{In the rest of the paper we refer only to activities for simplicity, but all the concepts apply to both activities and fragments.}.
%\end{definition}
\vspace{-0.1cm}
%In the rest of the paper we mainly refer to activities, but all the concepts also apply to the fragments hosted in an activity.

Stop-start events are generated when the execution of an activity must be temporarily suspended. There are many common situations that produce stop-start events. For instance, answering a phone call requires suspending the execution of the app, and thus also of the current activity, until the call ends. Moving an app to the background may cause its activities to be destroyed. When the app is moved again to the foreground, the status of the foreground activity has to be recreated. The rotation of the screen finally causes the destruction of the current activity that must be redrawn with a new layout. 

Note that all these situations are neutral from the user's perspective, that is, they are not expected to change the status of the app: users expect to find the status of an app unchanged after they have answered a phone call, after the app has been moved to the background and then to the foreground, and after the screen has been rotated. However, this behavior is not provided for free by Android, but it must be guaranteed by developers who have to implement the logic to save and retrieve the status of the activities. If this piece of logic is not implemented correctly, stop-start events are no longer neutral and state information might be lost, causing a \emph{data loss} problem.

\vspace{-0.1cm}
\begin{definition}
A \emph{data loss} problem occurs when data is accidentally deleted or state variables are accidentally assigned with default or initial values.   
\end{definition}
\vspace{-0.1cm}

Data loss faults can be the source of diverse failures~\cite{Jha:AndroidCrash:Mobilesoft:2019}. Indeed, the initialization of some program variables with wrong values (e.g., to the default value) %determined by their types) 
can be the cause of unpredictable behaviors. Based on our evaluation, we isolated five main failure patterns which are exemplified in Figure~\ref{fig:crash}: %(in the visual examples, the data loss faults are exercised by rotating the screen of the apps).

%\noindent  \emph{Crashes}: the app may simply crash. This is exemplified in Figure~\ref{fig:crash} (a) where the \app{Easy xkcd} app crashes after a rotation of the screen.
%
%\noindent \emph{Destroyed GUI elements}: some graphical elements may disappear forcing the user to repeat operations. This is exemplified in Figure~\ref{fig:crash} (b) where the calendar dialog in the \app{Diary} app disappears after a rotation of the screen.
%
%\noindent \emph{Phantom GUI elements}: some graphical elements may erroneously appear, forcing the user to perform unwanted and unclear interactions. This is exemplified in Figure~\ref{fig:crash} (c) where a dialog appears in the \app{CycleStreets} app after a rotation of the screen.
%
%\noindent \emph{Modified values}: some elements may unexpectedly change their values, resulting in misbehaviors of the app. This is exemplified in Figure~\ref{fig:crash} (d) where multiple text fields change to 0 in the \app{BeeCount} app after a rotation of the screen.
%
%\noindent \emph{Compromised internal state}: the internal state of the app might be compromised causing visible misbehaviors in the interactions that follow the activation of the data loss. This is exemplified in Figure~\ref{fig:crash} (e). The initial rotations of the screen compromise the status of the \app{Taskbar} app without producing any visibile misbehavior, until the last rotation causes a crash.

\begin{itemize}[leftmargin=*]
\item \emph{crashes}: the app may simply crash. This is exemplified in Figure~\ref{fig:crash} (a) where the \app{Easy xkcd} app crashes after a rotation of the screen.
\item \emph{destroyed GUI elements}: some graphical elements may disappear forcing the user to repeat operations. This is exemplified in Figure~\ref{fig:crash} (b) where the calendar dialog in the \app{Diary} app disappears after a rotation of the screen.
\item \emph{phantom GUI elements}: some graphical elements may erroneously appear, forcing the user to perform unwanted and unclear interactions. This is exemplified in Figure~\ref{fig:crash} (c) where a dialog appears in the \app{CycleStreets} app after a rotation of the screen.
\item \emph{modified values}: some elements may unexpectedly change their values, resulting in misbehaviors of the app. This is exemplified in Figure~\ref{fig:crash} (d) where multiple text fields change to 0 in the \app{BeeCount} app after a rotation of the screen.
%some elements may unexpectedly change their values, as exemplified in Figure~\ref{fig:crash} (d) where multiple text fields change to 0 in the \app{BeeCount} app after a rotation of the screen.
\item \emph{compromised internal state}: the internal state of the app might be compromised causing visible misbehaviors in the interactions that follow the activation of the data loss. This is exemplified in Figure~\ref{fig:crash} (e). The initial rotations of the screen compromise the status of the \app{Taskbar} app without producing any visibile misbehavior, until the last rotation causes a crash. %. However, if the user scrolls and rotates again the app, the effect of the scrolling is lost, due to the corrupted state induced by the previous rotation.
\end{itemize}

\section{\fullname}\label{sec:approach}

\fullname addresses data loss faults combining three key ingredients: (i) a biased model-based exploration strategy, which increases the likelihood to explore states that may expose data loss faults; (ii) data-loss-revealing actions, which interact with the app under test stimulating behaviors that are prone to data loss, and (iii) data loss oracles, which analyze the behavior of the app under test to detect data loss failures. %We describe these three key elements in the rest of this section.

\subsection{Biased Model-Based Exploration} \label{sec:exploration}
%\begin{todo}
%- stato = activity (nome), lista eventi che puoi produrre (enabledeness abstraction)  ** Droidbot troppo dettagliato (servizi attivi, proprieta), scrivo label e produco doppia rotazione, quindi troppe rotazioni inutili (astraendo dall'identita della widget ma considerando solamente il suo tipo)
%
%- bias esplorazione 
%\end{todo}

The test case generation strategy implemented in \name consists in visiting as many states as possible and incrementally testing the newly discovered states to detect data loss faults. To this end, the strategy builds a GUI model that represents the visited states and the executed actions. The model serves two main purposes: to distinguish the already visited (and tested) states from the new ones, and to bias the exploration towards the execution of actions that may potentially lead to states never visited before.

%More formally, 
\begin{definition}
A \emph{GUI model} is a non-deterministic finite state automaton %(NFA) 
%\begin{center} 
($Q$, $\Sigma$, $q_0$, $\delta$), 
%\end{center}
where  $Q$ is a finite set of \textit{abstract states}; $\Sigma$ is the finite set of \textit{events} that can be triggered from such abstract states, such as clicks, swipes, or \textit{stop-start} events; $q_0 \in Q$ is the \textit{initial abstract state}; $\delta: Q \times \Sigma \rightarrow \wp(Q)$ is the \textit{transition function}, which, given $q \in Q$ and $e \in \Sigma$, returns the set of \textit{abstract states} reachable from $q$ by executing $e$.%; $F = Q$ is the set of \textit{final abstract states} (all the \textit{abstract states} are accepting states).

\end{definition}

%\begin{revise}Figure~\ref{fig: model} shows an excerpt of a sample GUI model extracted for the ?? application.\end{revise}

%\begin{figure}
%\centering
%\begin{tikzpicture}[shorten >=1pt,node distance=2cm,auto]
%  \tikzstyle{every state}=[fill={rgb:black,1;white,10}]
%
%  \node[state,initial,accepting](s){$\:\:\:Main\:\:$};
%  \node[state, accepting](q_1)[below left of=s]{$\:\:\:Login\:\:$};
%  \node[state, accepting](q_2)[below of=q_1]{$Settings$};
%  \node[state, accepting](r_1)[below right of=s]{$\:\:\:Menu\:\:$};
%  \node[state, accepting](r_2)[below of=r_1]{$\:\:\:Menu\:\:$};
%
%  \path[->]
%  (s)   edge [loop right] node {$e_1$, $e_2$} (s)
%        edge              node {$e_3$} (q_1)
%        
%  (q_1) edge [loop left]  node {$e_1$} (q_1)
%        edge [bend left]  node {$e_4$} (q_2)
%        
%  (q_2) edge              node {$e_2$} (r_2)
%        edge [bend left]  node {$e_3$} (q_1)
%        
%  (r_1) edge              node {$e_1$} (q_1)
%        edge              node {$e_2$} (s)
%        edge [loop right] node {$e_3$} (r_1)
%        
%  (r_2) edge              node {$e_2$} (q_1)
%        edge              node {$e_5$} (r_1);
%\end{tikzpicture}
%\caption{Example of the GUI model}
%\label{fig: model}
%\end{figure}

To effectively test an app, it is important to represent states at an appropriate abstraction level. A too abstract representation would collapse many concrete states into a single abstract state causing several relevant states not being tested for data loss. A too concrete representation would cause an enormous waste of time, testing for data loss states with irrelevant differences. The abstraction level used by \name derives from the following two observations.

%\emph{Concrete values do not matter}: a GUI state representation might include all the widgets with their properties and values. As a consequence, two concrete states that differ for a single value assigned to a label or to an input field would produce different abstract states that would be tested for data loss. This is a waste of resources, because as long as some values are assigned to the various GUI elements, the data loss is likely to show up regardless of the concrete values assigned to these elements.
%
%\emph{Totally ignoring the content of screens is too inaccurate}: a GUI state representation might be so abstract to only consider the name of the current Android activity as identifier of the abstract state, that is, the number of abstract states matches with the number of Android activities implemented in a tested app. This strategy totally ignores the content of the screens and is likely to miss all those data loss faults that can be exercised only in a specific state of an activity. %For instance, the \app{AntennaPod} 

\begin{itemize}[leftmargin=*]
\item \emph{Concrete values do not matter}: a GUI state representation might include all the widgets with their properties and values. As a consequence, two concrete states that differ for a single value assigned to a label or to an input field would produce different abstract states that would be tested for data loss. This is a waste of resources, because as long as some values are assigned to the various GUI elements, the data loss is likely to show up regardless of the concrete values assigned to these elements.
\item \emph{Totally ignoring the content of screens is too inaccurate}: a GUI state representation might be so abstract to only consider the name of the current Android activity as identifier of the abstract state, that is, the number of abstract states matches with the number of Android activities implemented in the tested app. This strategy totally ignores the content of the screens and is likely to miss all those data loss faults that can be exercised only in a specific state of an activity. %For instance, the \app{AntennaPod} app fails due to a data loss only if a podcast has been loaded, and the same activity does not fail if no podcast was loaded.      
\end{itemize}

\vfill
 
Based on these two observations, \name uses an abstraction that ignores the concrete values, but still discriminates the relevant distinct states associated with a same Android activity. To this end, it uses the \emph{enabledeness abstraction}, which has already been used in other contexts to distinguish the states of software applications~\cite{deCaso:2011,deCaso:2013}, and preliminary experienced in Quantum to reveal GUI interaction faults~\cite{Razieh:OraclesUserInteraction:ICST:2014}. In this case, the idea is to distinguish states not based on the content of the screen, but based on the set of actions (i.e., events) that are allowed. Intuitively, it is worth distinguishing states that enable a different set of behaviors for the software. For instance, two different values in an input field produce two different abstract states only if one of the two values enables operations that are otherwise disabled. %More formally, 

\vfill

\begin{definition}
An \textit{abstract state} $q \in Q$ associated with a concrete state $s$ is a pair $(a, E)$, where $a$ is the name of the current activity in $s$ and $E \subseteq \Sigma$ is the %an ordered 
set of events enabled in $s$.
\end{definition}

\vfill

%\begin{definition}
%Let $q = (a, E)$ and $q' = (a',E')$ be two \textit{abstract states}, where $E = \{e_1, e_2, ..., e_n\}$ and $E' = \{e'_1, e'_2, ..., e'_n\}$, then:
%\begin{center} 
%    $q \equiv q' \Leftrightarrow a = a'  \land e_i = e'_i \; \forall i \in \{1, ..., n\}$
%\end{center}
%\end{definition}

The test case generation strategy is biased towards the execution of new actions that may lead to new (abstract) states potentially affected by data loss. %  while traversing the (abstract) states of the app under test. 
% by \name systematically executes new actions 
%is random, but \emph{biased} by a factor that steers the testing process towards exploring new abstract states, based on the model built during testing. 
In particular, every time an action is executed, \name has a probability $\epsilon$ to choose an event at random, and a probability $1-\epsilon$ to choose an event that has never been executed in the current abstract state based on the available GUI model. %, and a probability $1-\epsilon$ to choose an event at random. 

\name can generate five types of actions during the exploration (four actions inherited from DroidBot plus a scroll action added to reach every view in a window), in addition to the data-loss-revealing actions described in next section:
\begin{itemize}[leftmargin=*]
    \item \textit{TouchEvent}, which executes a tap on an clickable view;
    \item \textit{LongTouchEvent}, which executes a long tap on a clickable view;
    \item \textit{SetTextEvent}, which writes text inside an editable view;
    \item \textit{KeyEvent}, which presses a navigation button (e.g. ``Home'').%(e.g. ``Back'' or ``Home'').
    \item \textit{ScrollEvent}, which executes a swipe on a scrollable view;
\end{itemize}

\name incrementally updates the GUI model after the execution of every event. The testing activity stops after a budget that can be expressed as a number of actions to be performed or as an amount of time to be allocated for testing. 

\vfill

\subsection{Data-Loss-Revealing Actions} \label{sec:dataLossrevealing}
%\begin{todo}
%2 actions
%- fill-all (stringa costante) e 
%azione doppia rotazione su ogni nuovo stato, poi fa scroll pre rivelare elementi nascosti. 
%- azione doppia rotazioni a caso
%\end{todo}

The exploration activity described in Section~\ref{sec:exploration} %, aimed at reaching the abstract states that might be affected by data loss problems, 
is combined with the execution of data-loss-revealing actions that have the objective to reveal data loss problems, if present. 

\name includes two data-loss-revealing actions: one is executed \emph{systematically every time a new abstract state is reached} (\emph{systematic data-loss-revealing action}), while the other is executed \emph{probabilistically at every abstract state} (\emph{probabilistic data-loss-revealing action}). 

\newpage
The systematic data-loss-revealing action is composed of 
the following sequence of steps:
%
%multiple steps that are executed sequentially as follows:
\begin{enumerate}[leftmargin=*]
\item \emph{fill-in}: \name interacts with all the input views (e.g., text field, combo box, check box) entering non-empty values different from the default values, \label{set:firstStep}
\item \emph{save state}: the current GUI state is saved to later check if any data loss occurred. The way the state is saved depends on the oracle strategy adopted (see Section~\ref{sec:oracle}), \label{set:secondStep}
\item \emph{double screen rotation}: the screen rotation is a stop-start event. It is executed twice to reach a state that should be exactly the same that was saved, if no data loss occurred,
\item \emph{check state}: the current GUI state is compared to the saved state to determine if any data loss occurred. The way the comparison is performed depends on the oracle strategy (see Section~\ref{sec:oracle}), \label{set:lastStep}
\item \emph{scroll down}: some Android activities may include visual elements that span over the size of the screen. To make sure the reached abstract state is fully tested for data loss, including the elements that might be outside the screen, 
if the \emph{check state} step has not revealed a data loss, \name executes a scroll down action that may make new elements appear. If this happens, a new abstract state might be reached, which would be again systematically tested for data loss.  
\end{enumerate}

The systematic data-loss-revealing action already guarantees an accurate validation of the state space, as defined by our abstraction strategy. However, there might be certain data loss faults that depend on internal state information that does not produce any difference in the GUI state of the app. %For instance, Figure~\ref{fig:stateLost} shows the window for setting a timer in the \app{AntennaPod} app. This window produces a data loss only if a podcast has been loaded before. The fact that a podcast was loaded does not result in any visible difference in the timer window, and thus the abstraction strategy cannot capture the difference between the state that exposes the data loss fault and the one that does not. To address these cases, when an already visited state is encountered, \name enables the probabilistic data-loss-revealing action that has the same probability as the others to be selected. The probabilistic data-loss-revealing action performs the sequence of events from step~\ref{set:secondStep} to~\ref{set:lastStep} of the systematic data-loss-revealing action.
For instance, the window for setting a timer in the \app{AntennaPod} app generates a data loss only if a podcast has been loaded before from another window. The fact that a podcast was loaded does not result in any visible difference in the timer window, and thus the abstraction strategy cannot capture the difference between the state that exposes the data loss fault and the one that does not. To address these cases, when an already visited state is encountered, \name enables the probabilistic data-loss-revealing action that has the same probability of the other actions to be selected. The probabilistic data-loss-revealing action performs the sequence of events from step~\ref{set:secondStep} to~\ref{set:lastStep} of the systematic data-loss-revealing action.
%
%enables the probabilistic data-loss-revealing action, which performs the sequence of events from step~\ref{set:secondStep} to~\ref{set:lastStep} of the systematic data-loss-revealing action and has the same probability as the others to be selected.% The probabilistic data-loss-revealing action performs the sequence of events from step~\ref{set:secondStep} to~\ref{set:lastStep} of the systematic data-loss-revealing action.

In a nutshell, the exploration works as follows. When a new abstract state is encountered, \name executes the systematic data-loss-revealing action. Otherwise, \name identifies the set of actions $A$ that can be executed on the current GUI state based on the five types of supported actions (see Section~\ref{sec:exploration}). Namely a subset of them, $A^+$, has already been executed based on the GUI model, while the others, $A^-$, have not been executed yet. \name executes with probability $\epsilon$ a random action, that is, an action in the set $A \cup \{\textit{probabilistic}\ \textit{data-loss-revealing}\ \textit{action}\}$, and with probability $1-\epsilon$ an action that has not been executed yet, that is, an action in $A^- \cup \{\textit{probabilistic}\ \textit{data-loss-revealing}\ \textit{action}\}$.

\subsection{Data Loss Oracles} \label{sec:oracle}
%\begin{todo}
%- oracoli: GUI snapshot, widgets properties
%\end{todo}

Data loss problems do not always cause crashes. On the contrary, apps can present a range of misbehaviors as discussed in Section~\ref{sec:dataloss}. \name uses oracles based on the fact that the operations that exercise the data loss faults are expected to be neutral, thus leaving the status of the app unchanged. As anticipated in Section~\ref{sec:dataLossrevealing}, the basic \name strategy is to collect the GUI state of the app before and after the execution of actions that may have triggered a data loss failure and compare the states to detect it.

\begin{figure}[hbt!]
\centering
\begin{subfigure}[b]{0.47\linewidth}
    \centering
        \includegraphics[width=\linewidth]{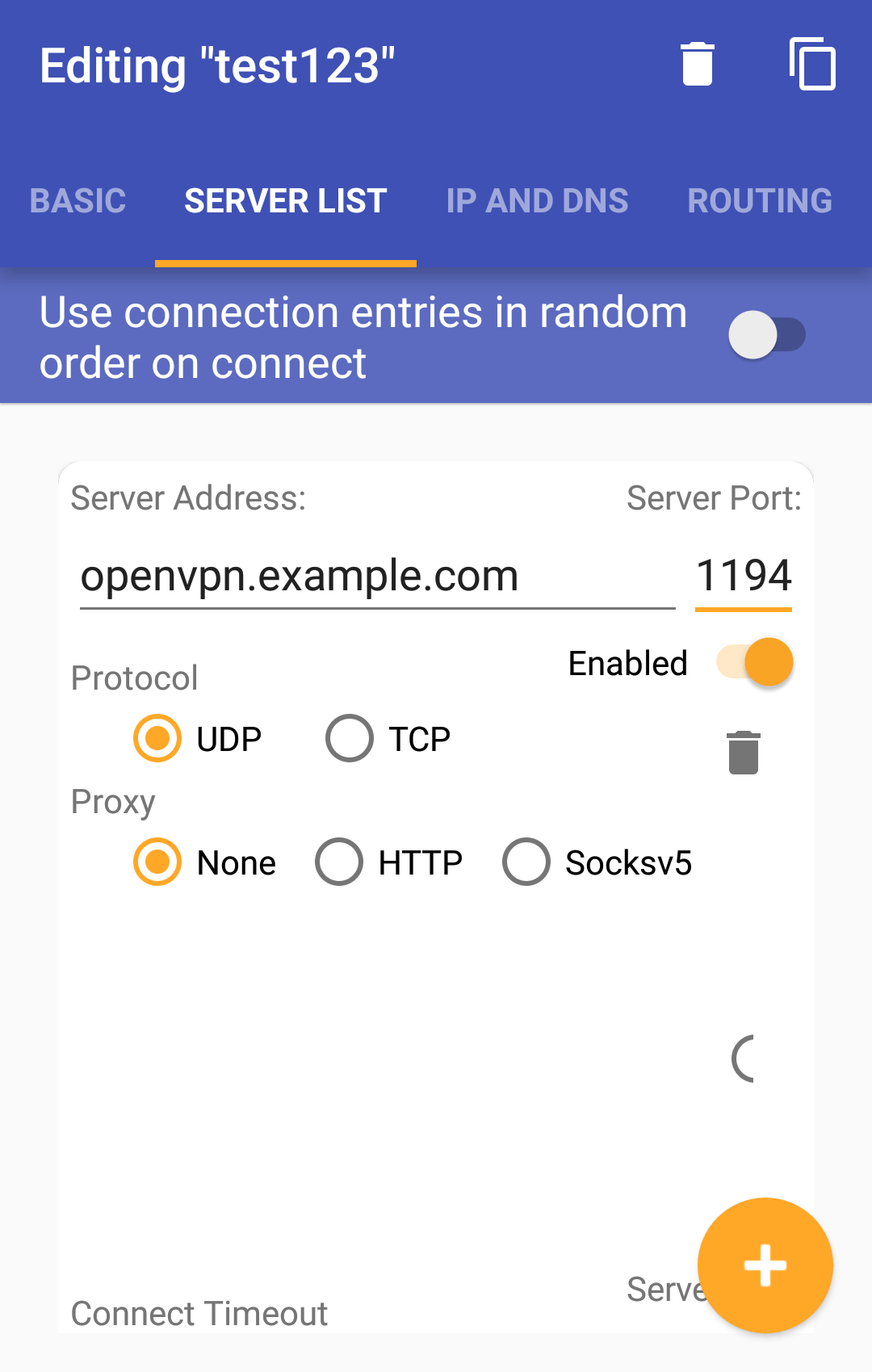}
        \caption{Snapshot-based oracle}
        \label{subfig:screeshotOracle}
    \end{subfigure}\hspace{3mm}%
%%   
%    \begin{subfigure}[b]{0.4\linewidth}
%    \centering
%    \end{subfigure} \hspace{3mm}%
%%
    \begin{subfigure}[b]{0.47\linewidth}
    \centering
	\begin{lstlisting}[language=json,frame=single]
[
  ...,
    {
      'content_description': None, 
      'resource_id': None, 
      'text': 'Editing "test123"', 
      'visible': True,
      'checkable': False, 
      'children': [],
      'size': '720*81',
      'checked': False,
      'temp_id': 4, 
      'selected': False,
      'child_count': 0,
      ...
   },
   ...
]
	\end{lstlisting}
	\caption{Property-based oracle}
	\label{subfig:propertyOracle}
	\end{subfigure}
\caption{The information saved by the two types of oracles for a same state of \app{OpenVPN}.}
\label{fig:oracles}
\end{figure}

%Similarly to other approaches~\cite{WhyDoesTheOrientationChangeMessUpMyAndroidApplication,Razieh:OraclesUserInteraction:ICST:2014}, 
\name defines two oracle strategies, which can be used either independently or jointly: the snapshot-based oracle and the property-based oracle. The \emph{snapshot-based oracle} takes a screenshot of the app before and after a data loss might have occurred and compares the images to detect failures. The \emph{property-based oracle} analyzes the GUI state and collects all the views and all their properties, and compares these two sets of properties to detect failures. Figure~\ref{fig:oracles} shows the state information captured by the snapshot-based and the property-based oracles for a same GUI state of the \app{OpenVPN} app. The former oracle stores a screenshot of the app, as shown in Figure~\ref{fig:oracles} (a), while the latter oracle stores the properties of the views in Python dictionary format, as shown in Figure~\ref{fig:oracles} (b).

More rigorously, the two oracle strategies collect and compare state information as follows.

\begin{description}[leftmargin=*]
\item[Snapshot-based oracle] \ \\%
	%\begin{itemize}
		 \emph{State Information}: \name first takes a screenshot of the device. The recorded image is then converted into a grayscale image, which is faster to compare than a colourful image. Finally, \name crops the header and the footer of the image because it contains information that changes over time regardless of data loss, such as the current time and the battery level. The resulting image is the retrieved representation of the current state.\\
		 \emph{State Comparison}: \name compares the two states by comparing the two corresponding images pixel by pixel.  Since a blinking cursor might cause a small level of noise in the representation of the images, the comparison fails only if more than 15 pixels every 10,000 pixels are different. 
		 %compares the two images pixel by pixel. Since a blinking cursor might cause a small level of noise in the representation of the images, the comparison fails only if more than 15 pixels every 10,000 pixels are different. 
	%\end{itemize}
\smallskip
\item[Property-based oracle]\ \\%
	%\begin{description}
		\emph{State Information}: \name retrieves all the views, including their properties and their hierarchical organization. The retrieved values are represented in a Python dictionary format.\\
		\emph{State Comparison}: \name compares the two states by comparing their Python-based representation. The comparison fails if one of the attribute values is different or the hierarchical structure of the views is not preserved.
	%\end{description}
\end{description}

\begin{figure}[hbt!]
	\centering
    \begin{subfigure}[t]{0.47\linewidth}
    	\centering
        \includegraphics[width=0.8\linewidth]{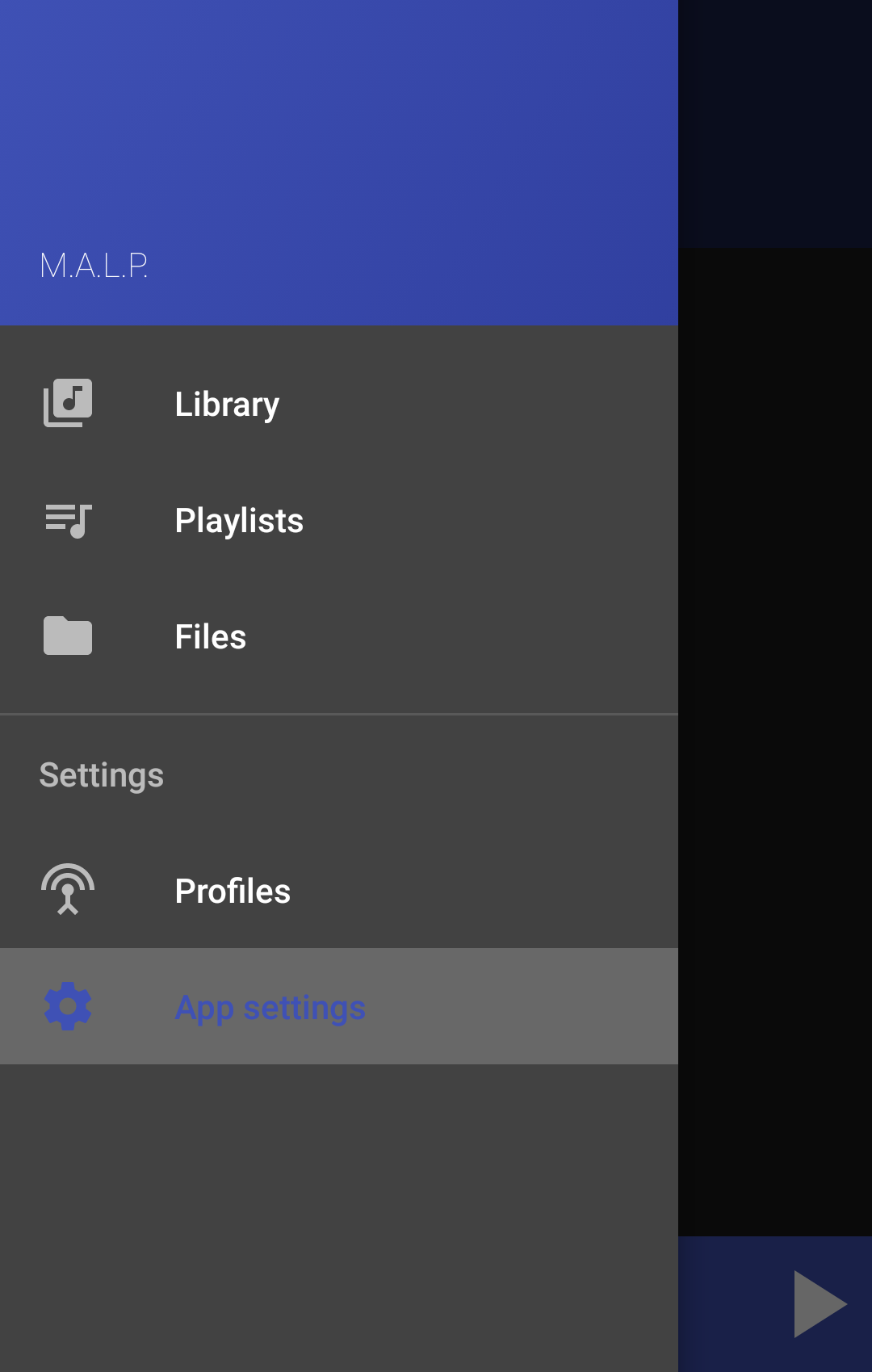}
%        \caption{Before double rotation}
    \end{subfigure} \hspace{3mm}%
    \begin{subfigure}[t]{0.47\linewidth}
    	\centering
    	\includegraphics[width=0.8\linewidth]{malp_dataloss_views.png}
  %      \caption{After double rotation}
    \end{subfigure}
  
	\begin{subfigure}[b]{0.47\linewidth}
	\begin{lstlisting}[language=json,frame=single]%,title={(a) Before a \textit{DoubleRotationEvent}},captionpos=b]	

[
  ...,
  {
    'content_description': 'Close navigation drawer',
    ...
  },
  ...,
]
	\end{lstlisting}
	\caption{Before double rotation}
    	\end{subfigure}\hspace{3mm}%
	 \begin{subfigure}[b]{0.47\linewidth}
	\begin{lstlisting}[language=json,frame=single]%,title={(b) After a \textit{DoubleRotationEvent}},captionpos=b]

[
  ...,
  {
    'content_description': 'Open navigation drawer',
    ...
  },
  ...,
]
\end{lstlisting}
\label{fig:view-based oracle}
\caption{After double rotation}
\end{subfigure}
\caption{A data loss failure detected by the property-based oracle only in \app{MALP} 3d31062.}
\label{fig:oracles-diff}
\end{figure}

\begin{figure}[hbt!]
   \centering
    \begin{subfigure}[t]{0.47\linewidth}
        \begin{center}\includegraphics[width=0.8\linewidth]{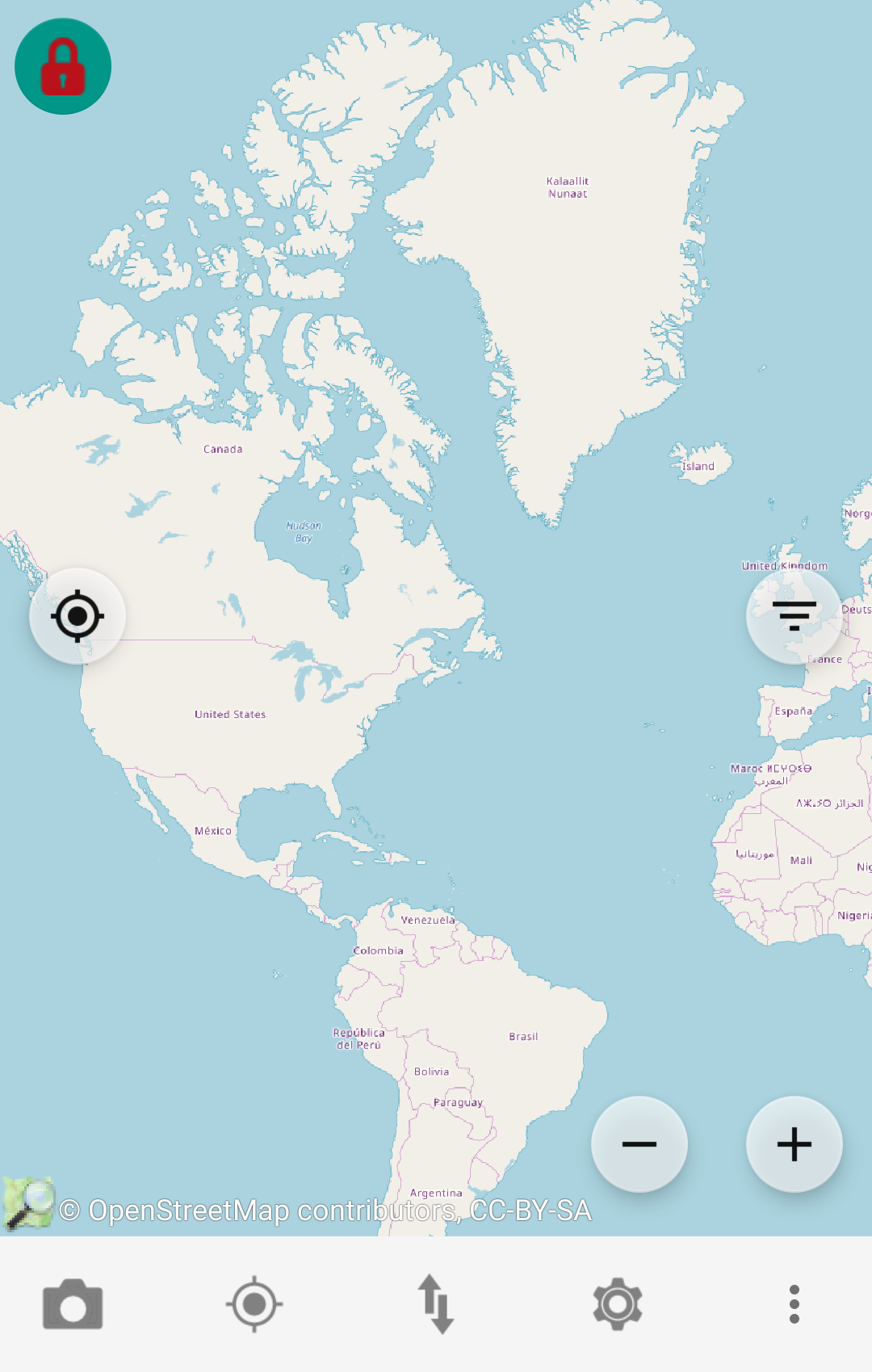}\end{center}
        	\caption{Before double rotation}
    \end{subfigure} \hspace{3mm}%
    \begin{subfigure}[t]{0.47\linewidth}
	\begin{center}\includegraphics[width=0.8\linewidth]{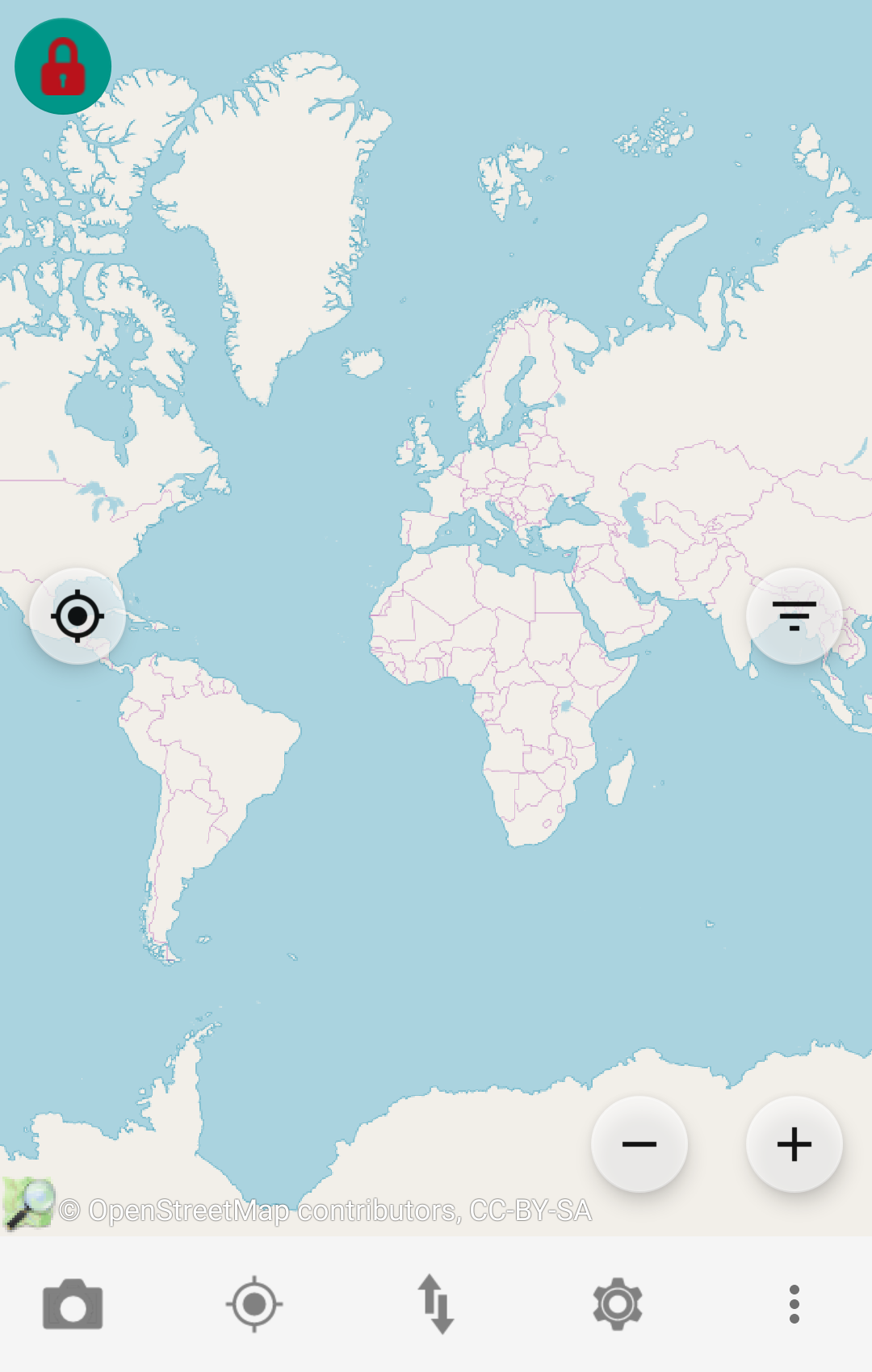}\end{center}
        	\caption{After double rotation}
    \end{subfigure}
\caption{A data loss failure detected by the snapshot-based oracle only in \app{Vespucci Osm Editor} v10.2.}
\label{fig:oracles-diff2}
\vspace{-0.5cm}
\end{figure}

Note that although the two strategies may seem redundant, they are not, as confirmed by our experiments. In particular, there are a number of data loss problems that can only be detected by one strategy. 
For instance, Figure~\ref{fig:oracles-diff} shows the case of a data loss failure that has been detected by the property-based oracle only. In fact, the two snapshots of the \app{MALP} app are visually identical, but actually the content descriptor has changed its value. On the other hand, Figure~\ref{fig:oracles-diff2} shows the case of a data loss failure that has only been detected by the snapshot-based oracle. The set of properties, not shown in the figure, are exactly the same for the two states, but the app has lost the zoom, as clearly visible from the snapshots. 

Interestingly, the two oracles can be combined, so that a failure is reported if just one of the two oracle strategies reports a failure. 

\section{Evaluation}\label{sec:evaluation}
This section presents the empirical results obtained by experimenting \name with a benchmark of \benchmarkfaults data loss faults, in comparison to the \alaric~\cite{alaric} and Quantum~\cite{Razieh:OraclesUserInteraction:ICST:2014} test case generation techniques. We first describe our implementation of \name, we then introduce the research questions and the subject applications. We finally present the results obtained for each research question in details, and discuss threats to validity.

%\smallskip
%\noindent \textbf{\large Implementation}
\subsection{Implementation} \label{sec:implementation}
%\subsection{Implementation} \label{sec:implementation}

%\noindent 
We implemented \name as an extension of \textit{DroidBot}~\cite{Droidbot}, which is a state-of-the-art test case generator for Android that does not implement features for the detection of data loss faults. \name inherits from DroidBot the capability to perform a specific sequence of actions before testing a target app. This feature can be used to setup the initial state of the app under test. In our evaluation, we used this capability to: authenticate into the apps that require a log in, setup an initial project in MGit, and grant permissions in \app{QuickLyric}.  

As outcome of the testing process, \name generates both a report with the revealed data loss faults and reproducible test cases. Each data loss is described in terms of the screenshots and the set of GUI properties collected before and after the data loss is observed. \name is available at \toolurl.

%\smallskip
%\noindent \textbf{\large Research Questions} 
\subsection{Research Questions} 

%\noindent 

We evaluate \name by studying the following five research questions.

\begin{itemize}[leftmargin=*]

\item \emph{RQ0 - \rqa} This research question studies the impact of the $\epsilon$ parameter on the effectiveness of the exploration. The result of this research question is used to configure \name to address the other research questions.  

\item \emph{RQ1 - \rqb} This research question investigates the effectiveness of \name considering two perspectives captured by the following sub-RQs. 

\begin{itemize}
\item \emph{RQ1.1 - What is the data loss discovery capability of \name?}  
\item \emph{RQ1.2 - What is the rate of the spurious oracle violations reported by \name?}
\end{itemize}

\item \emph{RQ2 - \rqc} This research question is decomposed into the following sub-RQs. 
\begin{itemize}
\item \emph{RQ2.1 - What is the relative effectiveness of \name, \alaric, and Quantum?} This sub-RQ compares \name to \alaric and Quantum.
\item \emph{RQ2.2 - What are the main factors that determine the effectiveness of \name?} This sub-RQ investigates the factors that allow \name to reveal more data loss faults than competing techniques.
\end{itemize}
%The first sub-RQ compares \name to \alaric and Quantum: \emph{RQ2.1 - What is the relative effectiveness of \name, \alaric and Quantum?}. While the second sub-RQ investigates the factors that allow \name reveal more data loss faults than competing techniques: \emph{RQ2.2 - What are the main factors that determine the effectiveness of \name?}.

%question compares the effectiveness of \name to the effectiveness of  \alaric and Quantum. We consider both the data-loss discovery, originating the sub-research question 

\item \emph{RQ3 - \rqd} This research question investigates the tradeoff between the two oracle strategies, measuring the data loss faults revealed by the snapshot-based oracle only, by the property-based oracle only, and by both.

\item \emph{RQ4 - \rqe} This research question studies how app developers react to the presence of data loss problems in their apps.

\end{itemize}

%\smallskip
%\noindent \textbf{\large Subject Applications} 
\subsection{Subject Applications}

%\noindent 
In our empirical evaluation we used the benchmark by Riganelli \emph{et al.}~\cite{ABenchmarkOfDataLossBugsForAndroidApps}, which includes \benchmarkfaults data loss problems affecting a total of \benchmarkApps Android apps and \benchmarkAppReleases app releases. Each data loss fault is equipped with an Appium test case~\cite{Appium} that can be executed to reproduce the problem.
 % (56 considering all their versions) and \textit{Appium} \cite{Appium} test cases useful for replaying and understanding such problems, where 98 of them also implement \textit{JUnit} \cite{JUnit} oracles to automatically detect the corresponding failures. In addition, it contains further 58 data loss problems already reported to the developers, but not provided with the test cases necessary for their replication. 
All the experiments have been conducted with the Genymotion v3.0.2 Android emulator using an emulated Google Nexus 5 device equipped with Android 6.0 API 23 and 2 GB of RAM. %. , while the Linux machine had an Intel Core I5-560M processor, 8 GB of RAM and ran Ubuntu 18.04.
In the evaluation, we followed the practice of other studies~\cite{AnEmpiricalStudyOfAndroidTestGenerationToolsInIndustrialCases} performing three runs of 3 hours each per tested app for both \name and \alaric, for a total of 42 days of uninterrupted computation.

%The choice for the number of runs and for the amount of time to be allocated for each run has been made relying on the study conduced by \cite{AnEmpiricalStudyOfAndroidTestGenerationToolsInIndustrialCases}, who performed an experimentation work to evaluate the main state-of-the-art test case generation tools on industrial Android apps. In their study, they carried out their experiments allocating 3 runs of 3 hours for each application in order to ``compensate potential influence brought by randomness during testing''. 
%
%Conforming to their decisions, since the stopping criteria for the exploration strategy can only be specified in terms of number of events to be triggered and since the emulator took about 4.8 seconds to complete the generation of a single event, every application of the benchmark has been tested with 3 runs of 2250 events. 

%\smallskip
%\noindent \textbf{\large RQ0 - \rqa}
\subsection{RQ0 - \rqa} \label{sec:rqa}

%\noindent 
This research question investigates the impact of the $\epsilon$ parameter on the effectiveness of the approach. To perform this initial study we selected 3 apps from the benchmark. To cover the range of situations that might be faced with the full benchmark, we selected the apps with the smallest, medium, and highest number of activities, which are \app{Equate} (2 activities), \app{Calendar Notification} (13 activities), and \app{Twidere} (52 activities), respectively.

To assess the capability to explore the app and potentially reveal data loss problems we measured \emph{activity coverage}, which is the percentage of activities covered in a test session. We started with $\epsilon=0$, which consists of a strategy that always privileges the execution of actions that have not been executed before, based on the incrementally constructed GUI model. We then increased the $\epsilon$ parameter by 0.1 to study its impact on the results. We stopped with $\epsilon=0.2$, which produced a significant decrease on the effectiveness of the exploration. Results are summarized in Table~\ref{table:epsilon}. We finally selected $\epsilon=0.1$ to address the other research questions.

\begin{table}[ht]
\caption{Impact of the $\epsilon$ parameter on activity coverage.}
\label{table:epsilon}
\begin{small}
\begin{tabular}{c | c}
\textbf{$\epsilon$} & \textbf{Activity Coverage} \\
\hline

$\epsilon=0$ (always new actions) & 52\% \\
$\epsilon=0.1$ (random actions with probability 0.1) & 60\% \\
$\epsilon=0.2$ (random actions with probability 0.2) & 44\% \\

\end{tabular}
\end{small}
\end{table}

%\smallskip
%\noindent \textbf{\large RQ1 - \rqb}
\subsection{RQ1 - \rqb} \label{sec:rq1}

%Before launching their executions, the apps have been granted with the permissions for accessing to the protected resources of the device (e.g., contacts, files, or camera) needed for their correct functioning.

%The default version of \textit{DroidBot} already provided a mechanism for translating JSON files, which must be written following specific structures, to ordered event sequences. Thanks to this functionality, since many apps required to login, different JSON files containing information about usernames and passwords have been written to inject the specific event sequences in order to complete the authentication. 
%The applications that required JSON files are the following:
%\begin{itemize}
%    \item K-9 Mail v5.010, v5.207 and v5.401
%    \item Octodroid v4.0.3 and v4.2.0
%    \item Tusky for Mastodon v1.0.3
%    \item Twidere v3.7.3
%\end{itemize}

%\noindent 
This research question investigates the capability of \name to reveal the data loss problems in the benchmark. To answer this research question, we manually analyzed every report produced by \name to distinguish the actual data loss problems from the irrelevant spurious violations. In particular, we classified a reported data loss as spurious if one of the two following conditions holds: (i) the state of the app after the double screen rotation is taken too early, while the activity is still recreating, making the oracle to fail its check or (ii) the difference reported by the oracle cannot be considered a data loss (e.g., because the tested app shows the current time which obviously changes after the screen has been rotated twice). In the vast majority of the cases the reports were enough to classify data loss problems. We reproduced the problem in the unclear cases. % (i) The screenshot obtained after the \textit{double rotation event} is black, which indicates a problem of the emulator (i.e., emulator not responding) and not a problem of the tested app; (ii) The screenshot after the \textit{double rotation event} is taken while the device screen is still rotating, which simply indicates that the emulator has not finished to process the rotation command yet; (iii) the difference reported by the oracle cannot be considered a data loss (e.g., because the tested app shows the current time which obviously changes after the screen has been rotated twice). 
For this research question, we detected data loss problems using both the snapshot-based and the property-based oracles. We analyze their relative fault detection ability with RQ3.

%
%\begin{definition}
%A \textit{false positive} occurs in one of the following situations:
%    \begin{itemize}
%        \item The screenshot obtained after the \textit{double rotation event} is black, which indicates that the emulator was not responding;
%        \item The screenshot after the \textit{double rotation event} was taken while the device screen was still rotating, which indicates that the emulator had not finished to process the adb command yet;
%        \item Either the screenshots or the views before and after the \textit{double rotation event} are different, but such differences could not be considered as data loss failures (see~Figure \ref{fig:falsepositive}).
%    \end{itemize}
%\end{definition}
%
%\begin{figure}[hbt!]
%\centering
%\begin{subfigure}[h]{0.45\linewidth}
%        \includegraphics[width=\linewidth]{images/false_positive_before.png}
%        \caption{Before a \textit{double rotation event}}
%    \end{subfigure}
%\begin{subfigure}[h]{0.45\linewidth}
%        \includegraphics[width=\linewidth]{images/false_positive_after.png}
%        \caption{After a \textit{double rotation event}}
%    \end{subfigure}
%\caption{A \textit{false positive} detected in World Clock \& Weather v1.8.6}
%\label{fig:falsepositive}
%\end{figure}

%\begin{definition}
%A \textit{true positive} is a data loss failure detected by the technique that is a real manifestation of a data loss problem.
%\end{definition}

We checked each data loss problem revealed by \name, distinguishing if it is a \emph{benchmark data loss}, that is, a problem that is part of the benchmark we used; an \emph{online data loss}, that is, a problem already reported online that is not part of the benchmark (for all the apps in the benchmark we searched online for additional data loss faults, and we used the snapshots, the activity name and the fields reported to lose their values to determine if a discovered data loss matches with the online data loss), or a \emph{new data loss}, that is, an unknown data loss problem (to the best of our ability to search for reported problems). We refer to the union of the benchmark and online data loss faults as the \emph{known data loss faults}. 

We report the \emph{number of activities affected by a data loss problem} found by \name. They intuitively correspond to different faults and different fixes to be implemented in different activities. 
The only exception is with the known data loss faults. Since some of the data loss faults in the benchmark affect the same activity, we actually report the precise number of faults in the benchmark that have been revealed. Finally, when only spurious violations are detected for an activity, we report it as a \emph{spurious data loss}.

%
%Once the execution phase was completed, that is, all the apps were successfully tested, the obtained raw data have been processed and classified with the following criteria:
%\begin{enumerate}
%    \item Every data loss failure detected by the technique was labelled as \textit{true positive} or as \textit{false positive};
%    \item Every data loss failure labelled as \textit{true positive} was further labelled as:
%    \begin{itemize}
%        \item \textit{Benchmark} if it belonged to the benchmark;
%        \item \textit{Online} if it did not belong to the benchmark but it had already been reported to the developers;
%        \item \textit{New} otherwise.
%    \end{itemize}
%    \item Every data loss failure was grouped into the corresponding \textit{activity} from which it was detected, thus avoiding considering the same data loss failures more than once;
%    \item Every \textit{activity} was labelled as:
%        \begin{itemize}
%            \item \textit{False positive} if all its data loss failures had been labelled as \textit{false positives};
%            \item \textit{Known} if at least one of its data loss failures had been labelled as \textit{benchmark} or \textit{online};
%            \item \textit{New} otherwise.
%        \end{itemize}
%\end{enumerate}
%%In conclusion, the data processed with the aforementioned criteria have been further analyzed in order to figure out the strengths and weaknesses of the proposed technique resulting from the experiments conduced.

% !TEX root =  main.tex

\begin{table*}[h]
%\caption{Results of the experiments.}
\caption{Results for DLD, and comparison to \alaric.}
%\vspace{-0.5cm}
\label{table: results1}
	\resizebox{1.0\linewidth}{!}{%
	\centering
	\begin{tabular}{lcc | ccccccc | c c c c c}
%	\toprule
%\cmidrule{4-15}
& & & \multicolumn{7}{|c}{\textbf{\name}} & \multicolumn{4}{|c}{\textbf{\alaric}}\\
%&&&&&&&&&&&&&&\\
%\multirow{5}{*}{\parbox{1.8cm}{\textbf{App name}}} &  
%\multirow{5}{*}{\parbox{1.8cm}{\textbf{Release}}} &  
%\multirow{5}{*}{\parbox{1.8cm}{\textbf{\# Activities}}} &  
%\multirow{5}{*}{\parbox{1.8cm}{\textbf{Activity Coverage avg (total)}}} &  
%\multirow{5}{*}{\parbox{1.8cm}{\textbf{Activities with \\Data Loss}}} &  
%\multirow{5}{*}{\parbox{2.0cm}{\textbf{Data Loss in Benchmark avg \\(total)/existing}}} &  
%\multirow{5}{*}{\parbox{1.8cm}{\textbf{Online Data Loss avg (total)}}} & 
%\multirow{5}{*}{\parbox{1.8cm}{\textbf{New Data Loss avg (total)}}} &  
%\multirow{5}{*}{\parbox{1.8cm}{\textbf{Spurious Violations avg (total)}}} &
%\multirow{5}{*}{\parbox[c]{1.8 cm}{\textbf{Crashes}}} &
%\multirow{5}{*}{\parbox{1.8cm}{\textbf{Activities with Data Loss}}} &  
%\multirow{5}{*}{\parbox{2.0cm}{\textbf{Data Loss in Benchmark avg \\(total)/existing}}} &  
%\multirow{5}{*}{\parbox{1.8cm}{\textbf{Online Data Loss avg (total)}}} &  
%\multirow{5}{*}{\parbox{1.8cm}{\textbf{New Data Loss avg (total)}}} &
%\multirow{5}{*}{\parbox{1.8cm}{\textbf{Spurious Violations avg (total)}}} \\
%
\cmidrule{4-15}
\textbf{App name} &
\textbf{Release} &
\textbf{\# Activities} &
\begin{tabular}{@{}c@{}} \textbf{Activity Coverage} \\ \textbf{avg (total)} \end{tabular}  &
\begin{tabular}{@{}c@{}} \textbf{Activities with} \\ \textbf{Data Loss} \end{tabular}  &
\begin{tabular}{@{}c@{}} \textbf{Benchmark}  \\ \textbf{Data Loss} \\ \textbf{avg (total)/existing} \end{tabular}  &
\begin{tabular}{@{}c@{}} \textbf{Online}  \\ \textbf{Data Loss} \\ \textbf{avg (total)/existing} \end{tabular}  &
\begin{tabular}{@{}c@{}} \textbf{New}  \\ \textbf{Data Loss} \\ \textbf{avg (total)} \end{tabular}  &
\begin{tabular}{@{}c@{}} \textbf{Spurious}  \\ \textbf{Data Loss} \\ \textbf{avg (total)} \end{tabular}  &
\textbf{Crashes} &
\begin{tabular}{@{}c@{}} \textbf{Activities with} \\ \textbf{Data Loss} \end{tabular}  &
\begin{tabular}{@{}c@{}} \textbf{Benchmark}  \\ \textbf{Data Loss} \\ \textbf{avg (total)/existing} \end{tabular}  &
\begin{tabular}{@{}c@{}} \textbf{Online}  \\ \textbf{Data Loss} \\ \textbf{avg (total)/existing} \end{tabular}  &
\begin{tabular}{@{}c@{}} \textbf{New}  \\ \textbf{Data Loss} \\ \textbf{avg (total)} \end{tabular}  &
\begin{tabular}{@{}c@{}} \textbf{Spurious}  \\ \textbf{Data Loss} \\ \textbf{avg (total)} \end{tabular}  \\
%&\multirow{5}{*}{\parbox{1.8cm}{\textbf{This is a very long line of text}}} & 

\toprule

Amaze File Manager & v3.1.0-beta.1 & 4 	& 100\% (100\%) 
	&\bb{3} 		&\bb{3 (3)/5}	&\bb{2 (2)/2}	&\bb{1(1)}	& 0 (0)		& 1	       
	& 1			&1 (1)/5 		&0 (0)/2 		&0 (0) 		& 0 (0)   \\ \midrule

AntennaPod & v1.5.2.0 & 16 & 33\% (44\%)
	&\bb{5}		&\bb{5 (5)/7}	&\bb{2 (2)/11}	&\bb{3 (3)}	& 1 (1)		&1 
	&1 			&0 (0)/7 		&1 (1)/11 		& 0 (0) 		&\bb{0 (0)}   \\ \midrule
	
BeeCount & v2.4.7 & 8 & 96\% (100\%) 
	&\bb{7}		&\bb{1 (1)/3}	&\bb{1 (1)/5} 	&5 (5) 	& 1 (1) 		& 0 
	&5			&0 (0)/3		&0 (0)/5		&5 (5) 	& 1 (1)   \\ \midrule

BookCatalogue & v5.2.0-RC3a & 35 & 66\% (71\%) 
	& \bb{21} 		&  \bb{5 (6)/7} 	& - 			& \bb{12 (15)} 	& \bb{0 (0)} 	& 1
	& 7 			&  2 (2)/7 		& - 			& 4 (5) 		& 6 (7) \\ \midrule

Calendar Notification & v3.14.159 & 13 & 67\% (69\%) 
	& \bb{8}  	& \bb{3 (3)/3} 	& \bb{1 (1)/1} 	& \bb{4 (5)} 	& 0 (0) & 1 
	& 2  		& 1 (1)/3 		& 0 (0)/1 		& 1 (1) 		& 0 (0)  \\ \midrule

CycleStreets & v3.5 & 11 & 55\% (55\%) 
	& \bb{6} 	& 1 (1)/1 	& - 	& \bb{5 (5)} & 0 (0) & 0
	& 1 		& 1 (1)/1 	& - 	& 0 (0) & 0 (0) \\ \midrule

Diary & v1.26 & 3 & 100\% (100\%) 
	& \bb{3}  	& \bb{1 (2)/2} 	& - 	& \bb{2 (2)} 	& 0 (0) & 0 
	& 2  		& 1 (1)/2 		& - 	& 1 (1) 		& 0 (0) \\ \midrule

DNS66 & v0.3.3 & 5 & 100\% (100\%) 
	& \bb{3} 	& \bb{1 (1)/1} 	& - 	& \bb{2 (2)} 	& \bb{0 (0)} & 0 
	& 1 		& 0 (0)/1 	& - 	& 1 (1) 		& 2 (2) \\ \midrule

Document Viewer & v2.7.9 & 9 & 48\% (56\%) 
	& \bb{2} 	& 1 (1)/1 		& - 	& \bb{2 (2)} 	& \bb{1 (1)} & 0
	& 1		& 1 (1)/1 		& - 	& 0 (0) 		& 4 (4)\\ \midrule

Easy xkcd & v6.0.4 & 9 & 74\% (78\%) 
	& \bb{4} 	& \bb{1 (1)/1} 	& - 	& \bb{3 (3)} 	& \bb{0 (0)} & 3
	& 1 		& 0 (0)/1 		& - 	& 1 (1) 		& 1 (1) \\ \midrule

Equate & v1.6 & 2 & 100\% (100\%) 
	& \bb{2}  	& 2 (2)/2 	& 1 (1)/1 	& \bb{1 (1)} 	& 0 (0) & 1
	& 1  		& 2 (2)/2 	& 1 (1)/1 	& 0 (0) 		& 0 (0) \\ \midrule

Etar Calendar & v1.0.10 & 12 & 42\% (42\%) 
	& \bb{3}  	& \bb{4 (4)/5} 	& \bb{2 (3)/5} 	& \bb{1 (1)} & 0 (0) & 0 
	& 0  		& 0 (0)/5 		& 0 (0)/5 		& 0 (0) & 0 (0) \\ \midrule

Firefox Focus & v4.0 & 6 & 44\% (50\%) 
	& \bb{3} 	& 0 (0)/1 	& - 	& \bb{3 (3)} 	& 0 (0) & 0
	& 2 		& 0 (0)/1 	& - 	& 2 (2) 		& 0 (0) \\ \midrule

Flym & v1.3.4 & 6 & 83\% (83\%) 
	& \bb{4} 	& 0 (0)/1 	& - 	& \bb{4 (4)} 	& 0 (0) & 0
	& 3 		& 0 (0)/1 	& - 	& 3 (3) 		& 0 (0) \\ \midrule

Gadgetbridge & v0.25.1 & 20 & 28\% (30\%) 
	& \bb{4}  	& \bb{1 (1)/1} 	& - 	& \bb{2 (3)} 	& 2 (2) & 2
	& 0  		& 0 (0)/1 		& - 	& 0 (0) 		& \bb{0 (0)}\\ \midrule

KISS Launcher & v2.25.0 & 2 & 100\% (100\%) 
	& 1 & \bb{1 (1)/1} & - & 0 (0) & 1 (1) & 0 
	& 1 & 0 (0)/1 & - & \bb{1 (1)} & \bb{0 (0)} \\ \midrule

Loop Habit Tracker & v1.6.2 & 7 & 71\% (71\%)  
	& \bb{4} 	& \bb{2 (2)/6} 	& - 	& 1 (2) 		& \bb{1 (1)} & 0  
	& 2 		& 0 (0)/6 		& - 	& \bb{2 (2)} 	& 2 (2)   \\ \midrule

MALP & 3d31062 & 2 & 100\% (100\%) 
	& \bb{1} 	& \bb{1 (1)/1} 	& - 	& 0 (0) & 0 (0) & 1
	& 0 		& 0 (0)/1 		& - 	& 0 (0) & 0 (0) \\ \midrule

MALP & v1.1.0 & 4 & 33\% (50\%) 
	& \bb{2} & \bb{2 (3)/4} 	& - 	& 1 (1) & 0 (0) & 1
	& 1 & 0 (0)/4 		& - 	& 1 (1) & 0 (0) \\ \midrule
	
MTG Familiar & v3.5.5 & 2 & 50\% (50\%) 
	& 1 	& \bb{1 (1)/1} 	& - & 0 (0) & 0 (0) & 0  
	& 1 	& 0 (0)/1	 	& - & \bb{1 (1)} & 0 (0) \\ \midrule

Notepad & v2.3 & 3 & 67\% (67\%) 
	& \bb{2}  	& 1 (1)/1 	& - 	& \bb{1 (1)} 	& 0 (0) & 0 
	& 1  		& 1 (1)/1 	& - 	& 0 (0) 		& 0 (0) \\ \midrule

Omni Notes & v5.4.3 & 17 & 29\% (35\%) 
	& \bb{4} 	& 0 (0)/1 	& - 	& \bb{4 (4)} & \bb{0 (0)} & 0  
	& 1 		& 0 (0)/1 	& - 	& 1 (1) & 1 (1) \\ \midrule

OpenTasks & v1.1.13 & 9 & 78\% (78\%)  
	& \bb{7} 	& \bb{1 (1)/1} 	& - 	& \bb{6 (6)} 	& \bb{0 (0)} & 0
	& 3 		& 0 (0)/1	 	& - 	& 3 (3) 		& 1 (1)   \\ \midrule

OpenVPN for Android & v0.7.5 & 13 & 46\% (46\%)  
	& \bb{5} 	& \bb{1 (1)/1} 	& - 	& \bb{4 (4)} 	& \bb{0 (0)} & 1 
	& 4 		& 0 (0)/1 		& - 	& 3 (4) 		& 1 (1)  \\ \midrule	

PassAndroid & v3.3.3 & 14 & 36\% (36\%) 
	& \bb{4}  		& \bb{2 (3)/3} 	& \bb{8 (8)/8} 	& 1 (1) 		& \bb{0 (0)} & 1
	& 3  		& 1 (1)/3 		& 4 (4)/8 		& 1 (1) 	& 1 (1) \\ \midrule
	
Periodic Table & v1.1.1 & 3 & 100\% (100\%) 
	& \bb{3} 	& \bb{2 (2)/2} 	& - 	& \bb{1 (1)} & \bb{0 (0)} & 0 
	& 0  		& 0 (0)/2 		& - 	& 0 (0) & 2 (2) \\ \midrule
	
Port Knocker & v1.0.8 & 6 & 50\% (50\%) 
	& 2  		& \bb{2 (2)/3} 	& 0 (0)/1 	& 0 (0) & \bb{0 (0)} & 0  
	& 2  		& 0 (0)/3 		& 0 (0)/1 	& \bb{2 (2)} & 1 (1) \\ \midrule
	
Prayer Times & v3.6.6 & 22 & 32\% (36\%) 
	& \bb{8} 	& \bb{3 (6)/7} 	& \bb{1 (1)/2} 	& \bb{4 (5)} 	& 0 (0) & 1  
	& 5  		& 1 (1)/7 		& 0 (0)/2 		& 3 (4) 		& 0 (0) \\ \midrule

QuasselDroid & v0.11.5 & 5 & 40\% (40\%) 
	& \bb{2} 	& \bb{1 (1)/1} 	& - 	& \bb{1 (1)} 	& \bb{0 (0)} & 1  
	& 0  		& 0 (0)/1 		& - 	& 0 (0) 		& 1 (1) \\ \midrule

Simple Draw & v3.1.5 & 7 & 86\% (86\%)  
	& \bb{3} 	& 0 (0)/1 	& - 	& \bb{3 (3)} 	& \bb{0 (0)} & 0
	& 0 		& 0 (0)/1 	& - 	& 0 (0) 		& 1 (1) \\ \midrule

Simple File Manager & v2.6.0 & 8 & 88\% (88\%) 
	& \bb{5}  	& 1 (1)/1 	& - 	& \bb{3 (4)} 	& \bb{0 (0)} & 2 
	& 1  		& 1 (1)/1 	& - 	& 0 (0) 		& 1 (1)  \\ \midrule

Simple File Manager & v3.2.0 & 8 & 75\% (75\%) 
	& \bb{5}  	& 1 (1)/1 	& - 	& \bb{3 (4)} 	& 0 (0) & 1 
	& 1  		& 1 (1)/1 	& - 	& 0 (0) 		& 0 (0)  \\ \midrule

Simple Gallery & v1.50 & 11 & 48\% (55\%) 
	& \bb{5}  	& \bb{1 (2)/4} 	& \bb{1 (1)/7} 	& \bb{3 (3)} 	& \bb{0 (0)} & 0
	& 1  		& 1 (1)/4 		& 0 (0)/7 		& 0 (0) 		& 1 (1) \\ \midrule
	
Simple Solitaire & v2.0.1 & 7 & 93\% (100\%) 
	& \bb{2}  	& \bb{1 (1)/1} 	& - 	& 1 (1) 	& \bb{0 (0)} & 1
	& 1  		& 0 (0)/1 		& - 	& 1 (1) 	& 2 (2) \\ \midrule
	
Simpletask & v10.0.7 & 11 & 67\% (73\%) 
	& \bb{7} 	& 1 (1)/1 	& - 	& \bb{6 (6)} 	& 0 (0) & 1 
	& 4 		& 1 (1)/1 	& - 	& 3 (3) 		& 0 (0) \\ \midrule

Syncthing & v0.9.5 & 9 & 93\% (100\%)  
	& \bb{8} 	& \bb{3 (4)/5} 	& \bb{2 (2)/2} 	& \bb{4 (5)} 	& 1 (1) & 2 
	& 2 		& 1 (1)/5 		& 1 (1)/2 		& 0 (0) 		& \bb{0 (0)} \\ \midrule

Taskbar & v3.0.3 & 21 & 26\% (29\%) 
	& \bb{3} 	& \bb{2 (2)/2} 	& \bb{12 (13)/13} 	& \bb{2 (2)} 	& 1 (1) & 1 
	& 1 		& 1 (1)/2 		& 1 (1)/13 			& 0 (0) 		& \bb{0 (0)} \\ \midrule

Tasks Astrid To-Do List Clone & v6.0.6 & 45 & 19\% (27\%) 
	& \bb{10}  	& 0 (0)/1 	& - 	& \bb{7 (10)} 	& 1 (1) & 2 
	& 1  		& 0 (0)/1 	& - 	& 1 (1) 		& \bb{0 (0)}  \\ \midrule

Vespucci Osm Editor & v10.2 & 19 & 42\% (47\%) 
	& \bb{7}  	& 1 (1)/1 	& - 	& \bb{5 (6)} 	& 1 (1) & 0 
	& 5  		& 1 (1)/1 	& - 	& 4 (4) 		& \bb{0 (0)}  \\ \midrule

Vlille Checker & v4.4.0 & 6 & 67\% (67\%) 
	& \bb{3}  	& \bb{1 (1)/1} 	& - 	& \bb{2 (2)} 	& 0 (0) & 0  
	& 1  		& 0 (0)/1 		& - 	& 1 (1) 		& 0 (0) \\ \midrule

WiFiAnalyzer & 1.9.2 & 4 & 75\% (75\%)  
	& \bb{3} 	& \bb{3 (3)/3} 	& - 	& \bb{1 (1)} 	& 0 (0) & 0  
	& 0 		& 0 (0)/3 		& - 	& 0 (0) 		& 0 (0) \\ \midrule

World Clock \& Weather & v1.8.6 & 4 & 100\% (100\%) 
	& \bb{4}  	& 0 (0)/1 	& - 	& \bb{4 (4)} 	& \bb{0 (0)} & 0
	& 1  		& 0 (0)/1 	& - 	& 1 (1) 		& 1 (1) \\ \midrule

\textbf{TOTAL} &  & 
\textbf{\ActivityNoLogin} & 
\textbf{\CoveredActivityNoLoginAvg (\CoveredActivityNoLoginMax)} 
& \textbf{\faultyActivityFoundNoLogin} 	&  \textbf{\revealedBenchmarkfaultsNoLogin/\benchmarkfaultsNoLogin} 	& \textbf{\onlineFaultsRevealed/\onlineFaults} &  \textbf{\newFaultsRevealedNoLogin} & \textbf{\AvgSpuriousViolationNoLogin (\SpuriousViolationNoLogin)} & \textbf{\NumCrashesNoLogin} 
& \textbf{\faultyActivityFoundNoLoginAlaric} 	& \textbf{\revealedBenchmarkfaultsNoLoginAlaric/\benchmarkfaultsNoLogin}		& \textbf{\onlineFaultsRevealedAlaric/\onlineFaults}	& \textbf{\newFaultsRevealedAlaric}		& \textbf{\AvgSpuriousViolationAlaric (\SpuriousViolationAlaric)}

 \\ \bottomrule
\multicolumn{15}{c}{} \\	

% \toprule
\multicolumn{15}{l}{\textbf{Apps tested after initial setup actions}} \\
 \cmidrule[\heavyrulewidth]{1-10}
 
Conversations & v1.14.0 & 21 & 41\% (57\%) 
	&  6  & 0 (0)/1 & - & 5 (6) & 0 (0) & \multicolumn{1}{c}{1} 
	%& x  & x & x & x & x  
	\\  \cmidrule{1-10}
Conversations & v1.23.8 & 23 & 40\% (57\%) 
	& 10  & 1 (1)/1 & - & 6 (9) & 0 (0) & \multicolumn{1}{c}{0} 
	%& x  & x & x & x & x  
	\\  \cmidrule{1-10}
K-9 Mail & v5.010 & 28 & 41\% (46\%) & 8  & 0 (0)/1 & - & 6 (8) & 0 (0) & \multicolumn{1}{c}{1} 
	%& x  & x & x & x & x  
	\\  \cmidrule{1-10}
K-9 Mail & v5.207 & 27 & 51\% (63\%)  & 7 & 1 (1)/1 & - & 5 (7) & 0 (0) & \multicolumn{1}{c}{1} 
	%& x  & x & x & x & x  
	\\  \cmidrule{1-10}
K-9 Mail & v5.401 & 29 & 54\% (59\%) & 8  & 1 (1)/1 & - & 6 (7) & 0 (0) & \multicolumn{1}{c}{1} 
	%& x  & x & x & x & x  
	\\  \cmidrule{1-10}
Mgit & v1.5.0 & 10 & 97\% (100\%) & 9  & 1 (1)/1 & - & 7 (8) & 1 (1) & \multicolumn{1}{c}{2}  
	%& x  & x & x & x & x  
	\\  \cmidrule{1-10}
OctoDroid & v4.0.3 & 44 & 56\% (66\%) & 21 & 2 (2)/2 & - & 16 (19) & 0 (0) & \multicolumn{1}{c}{3} 
	%& x  & x & x & x & x  
	\\  \cmidrule{1-10}
OctoDroid & v4.2.0 & 46 & 44\% (57\%) & 22  & 1 (1)/1 & - & 17 (21) & 1 (2) & \multicolumn{1}{c}{0} 
	%& x  & x & x & x & x  
	\\  \cmidrule{1-10}
%QuasselDroid & v0.11.5 & 5 & 40\% (40\%) & 3  & 1 (1)/1 & - & 2 (2) & 0 (0) & \multicolumn{1}{c}{1}  
	%& x  & x & x & x & x  
%	\\  \cmidrule{1-10}
QuickLyric & v2.1 & 4 & 75\% (75\%)  & 3 & 1 (1)/1 & - & 2 (2) & 0 (0) & \multicolumn{1}{c}{0}  
	%& x  & x & x & x & x  
	\\\cmidrule{1-10}
SMS Backup Plus & v1.5.11-Beta18 & 7 & 14\% (14\%) & 1 & 1 (1)/1 & - & 0 (0) & 0 (0) & \multicolumn{1}{c}{0} 
	%& x  & x & x & x & x  
	\\\cmidrule{1-10}
Tusky for Mastodon & v1.0.3 & 12 & 78\% (83\%) & 8  & 1 (1)/1 & - & 7 (7) & 0 (0) & \multicolumn{1}{c}{3} 
	%& x  & x & x & x & x  
	\\\cmidrule{1-10}
Twidere & v3.7.3 & 52 & 14\% (17\%) & 6 & 0 (0)/1 & - & 6 (6) & 0 (0) & \multicolumn{1}{c}{0}  
	%& x  & x & x & x & x  
	\\\cmidrule{1-10}

\textbf{TOTAL (all apps)} & & 
\textbf{\ActivityTotal} & 
\textbf{\CoveredActivityAvg (\CoveredActivityMax)} 
& \textbf{\faultyActivityFound} 	&  \textbf{\revealedBenchmarkfaults/\benchmarkfaults} 	& \textbf{\onlineFaultsRevealed/\onlineFaults} &  \textbf{\newFaultsRevealed} & \textbf{\AvgSpuriousViolation (\SpuriousViolation)} & \multicolumn{1}{c}{\textbf{\NumCrashes}}
%& \textbf{68} 	& \textbf{22/96}		& \textbf{8/58}	& \textbf{1 (34)}		& \textbf{1 (1)}
\\ \cmidrule[\heavyrulewidth]{1-10}

\end{tabular}
}
\end{table*}

%\begin{figure}
%\centering
%        \includegraphics[width=0.35\textwidth]{images/boxplot.png}
%\caption{FP rate of DLD}
%\label{fig:fp_dld}
%\end{figure}

Table~\ref{table: results1} column \name (left part of the table) shows the results that \name obtained for all the app releases considered in the study. Since every row represents the outcome of three runs, when applicable, we report both average values and total values. Column \emph{\# Activities} indicates the total number of activities in each app. Column \emph{Activity Coverage avg (total)} reports the activity coverage achieved in average and in total in the three executions.
Column \emph{Activities with Data Loss} indicates the number of activities affected by at least a data loss revealed by \name. Column \emph{Benchmark Data Loss avg (total)/existing} indicates the average and total number of data loss faults that have been revealed by \name out of the ones present in the benchmark. For example in the \app{BookCatalogue} app, \name revealed 5 data loss faults of the benchmark on average, achieved a total of 6 data loss faults revealed across the three runs, out of a total of 7 data loss faults present in the benchmark. Similarly, column \emph{Online Data Loss avg (total)/existing} indicates the average and total number of data loss faults reported online revealed by \name. Column \emph{New Data Loss avg (total)} indicates the average and total number of previously unknown data loss faults revealed.
Column \emph{Spurious Data Loss avg (total)} indicates the average and total number of activities that originated spurious violations only. Column \emph{Crashes} reports the total number of activities that crashed due to data loss faults. 
The top part of the table lists the apps where no initial setup has been necessary, while the bottom part of the table lists the apps that have been addressed as discussed in Section~\ref{sec:implementation}.

\subsubsection{RQ1.1 - What is the data loss discovery capability of DLD?}
%\textbf{\emph{RQ1.1}} 
In terms of exploration, \name managed to visit \CoveredActivityMax of the activities, revealing \faultyActivityFound activities affected by data loss faults (\percentageFaultyActivityFound of the total number of activities). It detected \revealedBenchmarkfaults of the \benchmarkfaults faults in the benchmark (\percentageRevealedBenchmarkfaults), \onlineFaultsRevealed out of the \onlineFaults (\percentageOnlineFaultsRevealed) additional data loss faults that we found online, and revealed \newFaultsRevealed new data loss faults (an average of \newFaultsPerRelease new data loss faults revealed per app). In total, \name revealed \totalFaultsRevealed data loss problems in \benchmarkAppReleases app releases, demonstrating a significant capability to detect data loss problems. Note that we started the empirical investigation knowing that less than \benchmarkfaults activities were affected by data loss problems and we ended up discovering \faultyActivityFound activities affected by data loss problems. 
\CHANGED{%Although most of the data loss faults have been naturally revealed by the systematic data-loss-revealing action, 
Interestingly, the probabilistic data-loss-revealing action contributed revealing data loss in 158 activities already reported by the systematic action and in 17 activities not reported by the systematic data-loss-revealing action.} 
%This is an evidence that the probabilistic action can detect problems outside the scope of the systematic action.}
%require particular iterations with the app under test to be detected.}

We manually investigated the \totalMissedKnownFaults cases of known data loss faults that have not been revealed by \name (\missedBenchmarkfaults cases in the benchmark and \onlineFaultsMissed cases retrieved online). We isolated three main reasons why data loss problems have been missed.

\begin{itemize}[leftmargin=*]
    \item \textit{Low probability sequences}: revealing these data loss failures requires the generation of an event sequence that has a low probability to be generated, due to the length of the sequence and/or the large number of actions that can be generated at every step of the testing process.
    \item \textit{Environment setup}: the detection of these data loss problems requires a specific set up of the environment. For instance, a data loss affecting the \app{Document Viewer} app requires the presence of a document to be revealed. These faults could be potentially revealed with additional effort in the setup of the app under test.
        \item \textit{Unsupported actions}: these data loss failures are impossible to reveal with \name because they require the execution of operations that are outside the scope of \name. For instance, detecting one of the data loss faults in the \app{Tasks Astrid To-Do List Clone} app requires to temporarily exit from the app and take a picture with the device camera, which cannot be done with \name.
\end{itemize}

Overall, we found \LowProbabilityMissedFaults data loss problems that simply have low probability to be revealed, \EnvironmentMissedFaults data loss problems that require a proper environment setup, and \NoActionMissedFaults data loss problems that require a more extensive exploration ability to be revealed. Interestingly, several faults might be potentially revealed by just executing \name for a longer time, or by refining the \name exploration strategy, so that specific combinations of actions are generated. However, generating complex and long combinations of actions can be challenging.

Finally, only \NumCrashes out of \faultyActivityFound activities affected by data loss faults produced crashes, which confirms the need of specific oracles to deal with these problems. 

%\textbf{\emph{RQ1.2}}
\subsubsection{RQ1.2 - What is the rate of the spurious oracle violations reported by DLD?}
\name performed well in terms of spurious oracle violations:  it reported only 1 activity with spurious data loss only every 4 tested apps, which indicates that \name is precise and seldom annoys testers with false alarms.

\begin{figure}
\centering
        \includegraphics[width=0.3\textwidth]{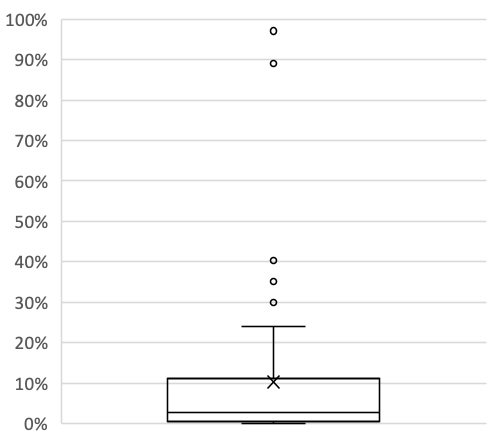}
\caption{Percentage of spurious oracle violations returned
per app.}%Percentage of detected data loss failures and spurious violations per oracle strategy.}
\label{fig: percentage_data-loss}
\end{figure}

Figure~\ref{fig: percentage_data-loss} shows the percentage of spurious oracle violations returned per app. The percentage ranges between a min of 0\% and a max of 24\%, with a mean value of 10.4\% and a median value of 2.7\%. Indeed, \name produces a limited percentage of spurious violations (less than 11\% for 75\% of the apps) that can be feasibly inspected by engineers when analyzing the output produced by the technique. The 5 outliers reported in Figure~5 correspond to apps with elements difficult to handle, such as timers and progress bars, that can be the source of an abnormal number of spurious violations due to spontaneous changes happening concurrently with the double rotations. We discuss the source of these spurious violations in RQ3.

\subsection{RQ2 - \rqc}

%\noindent 
This research question compares \name to both \alaric~\cite{alaric} and Quantum~\cite{Razieh:OraclesUserInteraction:ICST:2014}. \alaric represents the case of an \emph{alternative automated approach} to reveal data loss faults, while Quantum represents the case of an approach that can benefit from a \emph{manually generated model} to generate data loss-revealing test cases.

%\smallskip 

%\textbf{\emph{RQ2.1}} 
\subsubsection{RQ2.1 - What is the relative effectiveness of DLD, ALARic, and Quantum?}

The comparison to ALARic studies the effectiveness of the test generation strategy defined in \name, as described in Section~\ref{sec:approach}, to  \alaric, which uses a random (non-biased) exploration and a concrete states representation.

%We compare the test generation strategy defined in DLD to \alaric, which uses random (non-biased) exploration and a concrete states representation. 

We executed \alaric three times for 3 hours each time, as done for \name, and reported the results in Table~\ref{table: results1} (column \alaric). Note that we excluded from the comparison the apps that have been tested with \name exploiting an ad-hoc setup, since it would lead to an unfair comparison to \alaric, which does not implement this feature. %The numbers discussed below thus refer only to the apps tested with both approaches. 
Table~\ref{table: results1} shows with grey background the cases where a technique outperforms the other.

\name significantly outperformed \alaric in terms of data loss discovery capability. In fact, \alaric revealed \faultyActivityFoundNoLoginAlaric activities affected by data loss faults, while \name revealed \faultyActivityFoundNoLogin faulty activities, releasing a \faultDiscoveryDifference factor of improvement. \alaric found \revealedBenchmarkfaultsNoLoginAlaric data loss faults of the benchmark, \onlineFaultsRevealedAlaric online data loss faults, and \newFaultsRevealedAlaric new data loss faults. While \name revealed \revealedBenchmarkfaultsNoLogin of the data loss faults in the benchmark (\revealedBenchmarkfaultsNoLoginAlaricDLDgap improvement factor), \onlineFaultsRevealed online data loss faults (\onlineFaultsRevealedAlaricDLDGap improvement factor), and \newFaultsRevealedNoLogin new data loss (\newFaultsRevealedAlaricDLDGap improvement factor). Overall, \name revealed \emph{significantly more} data loss faults than \alaric. %Finally, note that the higher effectiveness of \name is confirmed on every single app, with all the faults revealed by \alaric also revealed by \name.%, where \alaric outperformed \name only once and limitedly to new data loss.   

\name performed better than \alaric also in terms of spurious data loss. In fact, \name produced spurious data loss only for \SpuriousViolationNoLogin activities, while \alaric produced spurious data loss for \SpuriousViolationAlaric of the activities. %Although generating few spurious data loss is desirable, \name still performs well with spurious data loss (only 1 every 4 apps). % and generating less spurious violations can be hardly traded with the detection of about one third of the faults. 

In summary, \name has been significantly more effective than \alaric with the studied apps.

%\smallskip

Since Quantum is not publicly available, we could not compare Quantum to \name on our set of apps. We thus executed \name for 3 hours on the same apps used in the evaluation of Quantum~\cite{Razieh:OraclesUserInteraction:ICST:2014} and compared the results. We limited the experiment to 4 of the 6 apps used to evaluate Quantum since for 2 apps it was impossible to retrieve the same version used in the original study. 

Since we do not know the manual effort that was necessary to manually define the models used by Quantum, it is hard to setup a fair comparison among the two approaches. However, the obtained results can still offer useful insights about the relative effectiveness of the two approaches.

% !TEX root =  main.tex

\begin{table}[ht]
%\vspace{-0.2cm}
\caption{Comparison between Quantum and \name.}
%\vspace{-0.4cm}
\label{table:quantum}
\begin{small}
\begin{tabular}{c | c | c | c | c}
 & \multicolumn{2}{c|}{\textbf{Quantum}} & \multicolumn{2}{c}{\textbf{\name}}\\
\textbf{App (Version)} & \multicolumn{2}{c|}{(with manual model)} & \multicolumn{2}{c}{(automatic)}\\
& \textbf{data} & \textbf{spurious} &\textbf{data} &   \textbf{spurious} \\
& \textbf{loss} & \textbf{violation} &\textbf{ loss} &   \textbf{violation} \\
\hline
%false positives
OpenSudoku (1.1.5)  &3 &2 &5&1\\

Nexes Manager (2.1.8)&7&2&11&1\\

VuDroid (1.4)&2&0&2&0\\

K9Mail (4.317)&4&1&15&2\\

\end{tabular}
\end{small}
%\vspace{-0.2cm}
\end{table}

Table~\ref{table:quantum} shows the distinct data loss and spurious violations reported by Quantum and \name. Although \name cannot benefit from a manual model, its activity has been more effective in revealing a number of data loss faults compared to Quantum. Of course, we do not know if the data loss faults revealed by \name include all the data loss faults revealed by Quantum. It might be the case that \name cannot reach some areas of the app under test that can be reached with the manual model. However, the results suggest that \name is quite effective even compared to techniques exploiting manual models.

%\textbf{\emph{RQ2.2}} 
\subsubsection{RQ2.2-What are the main factors that determine the effectiveness of DLD?}
To investigate the reason of the difference in the performance between \name and \alaric, we studied the reason why \alaric failed to reveal faults revealed by DLD. In particular, we counted 
%
%We studied the reason for the difference in the performance between \name and \alaric by counting 
the number of faulty activities not reached by \alaric and the number of faulty activities reached by \alaric without revealing the data loss fault. This produced the following results:
\begin{itemize}[leftmargin=*]
\item \alaric does not reach 52\% of the faulty activities revealed by \name only, that is, approximatively half of the additional data loss faults revealed by \name are due to a better exploration strategy,
\item \alaric reaches the faulty activity without revealing the data loss for 48\% of the faulty activities revealed by \name only. In a nutshell, the systematic fine-grained testing of the states implemented in DLD is more effective than Alaric's strategy,
\item 12 of the faulty activities missed by Alaric require a snapshot-based oracle to be revealed, but only 4 of them are reached by \alaric.
\end{itemize}

%\smallskip
%\noindent \textbf{\large RQ3 - \rqd}
\subsection{RQ3 - \rqd}

This research question investigates the complementarity between the snapshot-based and the property-based oracles. We already discussed in Section~\ref{sec:oracle} the qualitative differences between these two types of oracles, and reported examples of data loss faults that could be detected by one type of oracle only. Here, we assess quantitatively the impact of each class of oracles, on both the revealed data loss faults and the spurious oracle violations. 

\begin{figure}[hbt!]
\centering
        \includegraphics[width=0.44\textwidth]{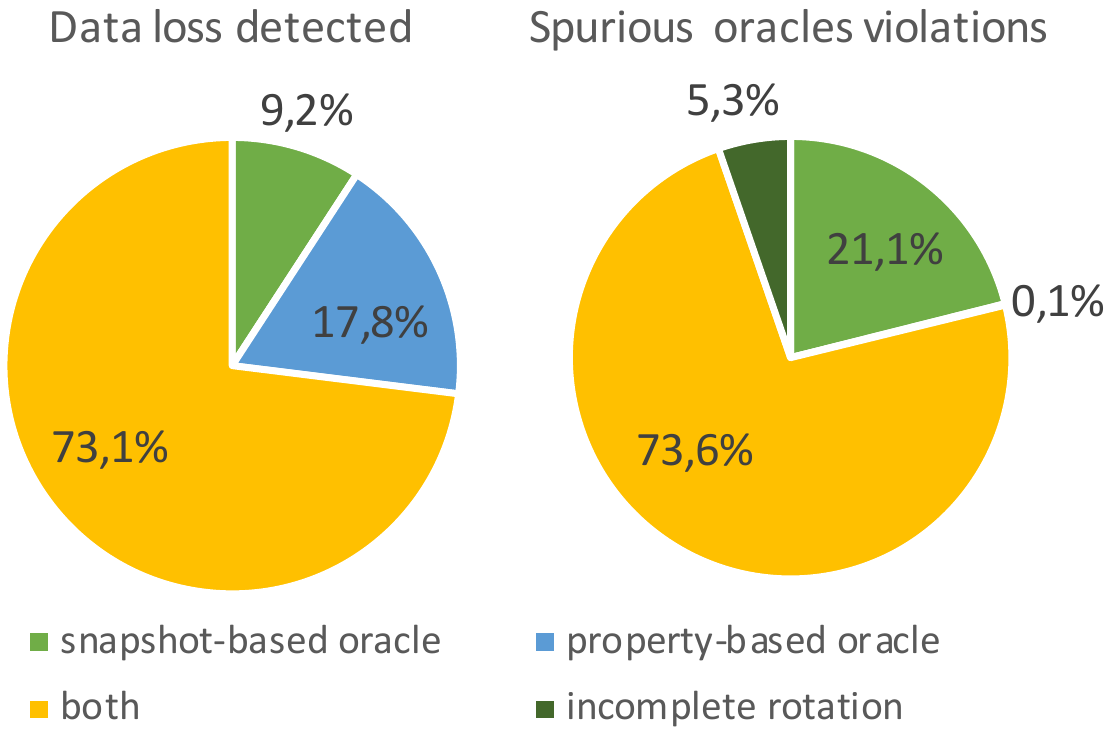}
\caption{Percentage of detected data loss failures and spurious violations per oracle strategy.}
\label{fig: oracles_total}
\end{figure}

Figure~\ref{fig: oracles_total} shows the percentage of data loss faults detected and the number of spurious oracle violations produced by one-strategy only, either snapshot-based or property-based, or both of them. In the case of the spurious violations we have an additional category that is the violations caused by slow activity recreation after screen rotation, as anticipated in the Section discussing RQ1. %~\ref{sec:rq1}.

None of the two approaches have been able to reveal every data loss problem. A large proportion of the failures (\dataLossDetectedByBoth) have been detected by both oracles, which implies that most of the data loss faults cause both properties that lose their values and visible issues on the app. However, there are yet \dataLossDetectedByOneOracle of the faults that require a specific type of oracle to be revealed.

In terms of absolute failure discovery ability, both oracles have been effective, with the property-based and snapshot-based oracles revealing \dataLossDetectedByPropertyOracle and \dataLossDetectedBySnapshotOracle of the failures, respectively.

A small number of the spurious oracle violations (\spuriousRotationOracle) is caused by a slow activity recreation, which causes the oracles to retrieve incorrect state information. This percentage can be reduced or eliminated by carefully tuning the timing of the oracles. 

Interestingly, the property-based oracle is also more effective in terms of spurious violations reported. In fact, only \spuriousViolationsByPropertyOracle of the spurious violations are produced uniquely by the property-based oracle, while \spuriousViolationsBySnapshotOracle of the spurious violations are produced uniquely by the snapshot-based oracle. The largest proportion of the spurious violations (\spuriousViolationsByBoth) are produced by both the strategies. 

Although the snapshot-based oracle produces more spurious violations than the property-based oracle, these violations seldom cause correct activities to be reported to the tester (1 activity every 4 apps), thus it is relatively detrimental to use it in the testing process. On the contrary, including it in the analysis increases the number of revealed data loss faults by \dataLossDetectedBySnapshotOracleOnly, which is a non-trivial increase of the failure discovery ability of \name.

%\begin{figure}[hbt!]
%\centering
%        \includegraphics[width=0.35\textwidth]{images/false_positives_oracles_piechart.png}
%\caption{Percentage of {Data Loss} failures labelled as \textit{false positives} detected by every oracle}
%\label{fig: false_positives_oracles_piechart}
%\end{figure}
%\begin{figure}[hbt!]
%\centering
%        \includegraphics[width=0.47\textwidth]{images/false_positives_oracles.png}
%\caption{Number of the data loss failures labelled as \textit{false positives} detected for each app}
%\label{fig: false_positives_oracles}
%\end{figure}

%\smallskip
%\noindent \textbf{\large RQ4 - \rqe}
\subsection{RQ4 - \rqe}

Finally, we investigated if data loss faults are relevant to app developers. To this end, for each app in the benchmark, we identified and downloaded the latest version of the same app. We executed again \name for 9 hours (three 3-hours runs) on each app and revealed 195 data loss faults that still affect these apps nowadays. We finally submitted a bug report online for each revealed data loss. 

Up to the time of the conference deadline, we received feedback for \feedbackReceived of the reports submitted online. Developers confirmed the bugs for \bugReportsConfirmed reports (\percentageBugReportsConfirmed of the reports with a feedback). Only in \bugReportsNotConfirmed cases developers rejected the report advocating that the fault was
a framework fault, claiming the fault was not reproducible, or giving no explanation. We can thus conclude that the revealed data loss faults are significant to the developers.

In \bugReportsNotRelevant of \bugReportsConfirmed cases developers claimed that the cost of fixing these bugs might be too high compared to their impact on the app. This decision of course depends on the specific consequence of the data loss and the complexity of the activity that is affected by the fault. This also suggests that the definition of an automatic repair strategy~\cite{Gazzola:Repair:TSE:2017} that can address data loss faults could be extremely beneficial to improve the cost-effectiveness of the bug fixing process.

%\smallskip
%\noindent \textbf{\large Threats to Validity}
\subsection{Threats to Validity}

The main internal threats to validity about our study is the manual work done to identify the spurious violations among the data loss faults reported in the evaluation. Distinguishing a genuine data loss from a spurious one is however quite simple, as also confirmed by the bug reports submitted online to developers that have been almost all accepted, with rejections due to faults that were considered outside the boundary of the tested app or behaviors that could be considered acceptable although not ideal.          

The main external threats to validity concerns with the generalization of the results. The significant number of faults and apps considered mitigates this threat. The fact that we repeated the evaluation with the most recent versions of the apps revealing again many data loss faults is a further mitigation factor. 

Concerning the comparison between \name and \alaric, the consistency of the results across every app that has been tested is a strong factor in favour of the generality of the results. The results of the comparison between \name and Quantum cannot be generalized due to the limited size and the setup of the experiment, however they still provide useful insights about the effectiveness of \name.

\section{Related Work}\label{sec:related} 

Several techniques covering a wide range of approaches are available to generate test cases for Android apps. 
For instance, Monkey generates test inputs fully randomly without interpreting the GUI of the app under test~\cite{Monkey}. A$^3$E systematically generates test inputs following a depth-first strategy~\cite{A3E}. DroidBot~\cite{Droidbot} and Stoat~\cite{Stoat} build a state-based model of the system under test and generate test cases exploiting information about the events already tested in the visited states. Sapienz uses evolutionary algorithms to generate test cases~\cite{Sapienz}. While these approaches revealed several interesting faults in both open source~\cite{AutomatedTestInputGenerationForAndroid:AreWeThereYet} and industrial applications~\cite{AnEmpiricalStudyOfAndroidTestGenerationToolsInIndustrialCases}, they are ineffective against data loss problems (and also against most non-crashing failures~\cite{Razieh:OraclesUserInteraction:ICST:2014}). In fact, they neither include operations that cause stop-start events nor they are equipped with oracles that can detect non-crashing data loss failures, which account for the majority of the failures as reported in our evaluation. 

Thor~\cite{SystematicExecutionOfAndroidTestSuitesInAdverseConditions} can augment existing test suites with neutral sequences of operations to reveal additional failures. The injected sequences concern with the audio service, the connectivity, and the lifecycle of the activities, which may reveal data loss faults. Similarly, when a user-generated model of the app under test is available, Quantum~\cite{Razieh:OraclesUserInteraction:ICST:2014} can generate tests that may reveal data loss, as reported in a small-scale evaluation. Differently from these approaches, \name directly generates the test cases and requires neither an initial test suite nor a model of the app under test, retaining a high effectiveness as demonstrated in the comparison to \alaric and Quantum. 

CrashScope~\cite{CrashScope} and AppDoctor~\cite{AppDoctor} can generate tests that may reveal data loss problems, but their effectiveness is limited to crashing faults, which represent the minority of the cases. 

\alaric~\cite{WhyDoesTheOrientationChangeMessUpMyAndroidApplication} is a mostly random test case generation technique that similarly to \name exploits double screen rotations to reveal data loss faults. However, the biased exploration strategy, the enabledeness state abstraction, and the ad-hoc data-loss-revealing actions used by \name outperformed \alaric in our evaluation.
%In fact, \name revealed three times more data loss faults than \alaric at the cost of a small increase in the spurious data loss reported. % (\name generated a spurious violation every 4 tested apps, in average).

%When a user-generated model of the app under test is available, Quantum can also be used to generate tests that may reveal data loss problems, as reported in a preliminary evaluation with 6 apps. Differently, \name is a fully automatic approach whose effectiveness has been studied with a large benchmark.

When the source code of the app is available, a static analysis technique such as KREfinder~\cite{Shan:ResumeRestart:OOPSLA:2016} can be used to reveal data loss problems. However, as most static analysis techniques, KREfinder suffers scalability issues and is likely to report many spurious violations. On the contrary, the effectiveness of \name does not depend on the complexity and size of the source code and seldom reports spurious data loss.

The oracle strategies presented in this paper relate to the work on metamorphic testing~\cite{Segura:metamorphic:TSE:2016}. Metamorphic testing exploits metamorphic relations, which are relations on multiple executions of the software, to check the correctness of the observed behavior. The neutral sequences of operations that \name uses to reveal data loss problems can be seen as a specific class of metamorphic relations that relate executions with and without these sequences. 

Relations between executions, like the ones used in this paper to reveal data loss problems, have been also used to heal executions, for instance to produce automatic workarounds~\cite{Carzaniga:workarounds:TOSEM:2015}. Although neutral sequences of events can be potentially used to obtain workarounds, the ones used in this paper can be hardly used to heal executions since they are often the cause of faults, as reported in the evaluation.

Finally, faults in Android apps, including data loss faults, could be addressed with healing techniques. However, not many healing approaches can work in the Android environment. 
Azim \textit{et al.} defined a technique that can disable functionalities that are not working properly~\cite{Azim:HealingSmartphones:ASE:2014}. While this approach might be exploited to prevent data loss failures, it also reduces the set of functionalities available to users. DataLossHealer is a healing solution designed to mitigate the impact of data loss faults in the field~\cite{HealingDataLossProblemsInAndroidApps}. Although it might prevent some data loss faults, it has various drawbacks. For instance, it introduces overhead to save and restore data in presence of data loss faults, it can address only some data loss, and it requires rooting the device. \name delivers a more effective solution revealing data loss faults upfront before the app is released.

% !TEX root =  main.tex
\vfill
\section{Conclusions}\label{sec:conclusion}

Android apps must be designed to deal with stop-start events, which are external events that may interrupt the execution of the running activity. %These events are generated in a number of common situations, for instance when answering phone calls and rotating the screen of the smartphone. 
When one of these events is generated, the foreground Android activity might be destroyed and later recreated. To avoid losing useful data during this process, apps must explicitly implement the logic necessary to save the data, when the activity is destroyed, and restore the saved data, when the activity is recreated. Unfortunately, this logic is often faulty~\cite{ABenchmarkOfDataLossBugsForAndroidApps,SystematicExecutionOfAndroidTestSuitesInAdverseConditions,WhyDoesTheOrientationChangeMessUpMyAndroidApplication,alaric}. 

%Despite their diffusion, test case generation techniques for mobile apps seldom consider the faults caused by stop-start events. 
This paper presents \fullname (\name), an automatic test case generation technique designed to reveal  data loss faults. \name exploits an exploration strategy biased towards the discovery of new app states, data-loss-revealing actions, and two dedicated oracle-based strategies to automatically reveal data loss problems. 

In our evaluation with \benchmarkfaults data loss faults affecting \benchmarkAppReleases app releases, \name outperformed \alaric~\cite{alaric} and performed well in comparison to Quantum when instructed with a manual model of the app under test. Overall, \name revealed \faultyActivityFound activities affected by data loss faults, which is a clear indicator of the effectiveness of the approach and pervasiveness of the problem. %The evaluation of the two oracle-based strategies, one based on snapshots and the other based on GUI properties, provided evidence of the complementarity between the two approaches. Finally, the bug reports submitted online starting from the revealed faults confirmed that the detected faults are actual faults.

We are now working on the definition of automatic program repair solutions for data loss faults. % that may suggest fixes that could be cost-effectively applied. 

\vfill

\smallskip
%\textbf{Acknowledgments}
\section*{Acknowledgments}\label{sec:acknowledgments}
We would like to thank Vincenzo Riccio, Domenico Amalfitano and Anna Rita Fasolino for sharing their implementation of \alaric with us.

\newpage

\balance
\bibliographystyle{ACM-Reference-Format}
  \bibliography{biblio}
\end{document}